\documentclass[10pt,a4paper]{article} 
\usepackage{graphicx}
\usepackage{style}
\usepackage{epstopdf}
\usepackage{epsfig}
\usepackage{amsmath,amssymb,bm,amsfonts,yfonts,bbm}

\newcommand{\be}{\begin{equation}}
\newcommand{\ee}{\end{equation}}
\newcommand{\bea}{\begin{eqnarray}}
\newcommand{\nn}{\nonumber}
\newcommand{\eea}{\end{eqnarray}}

\newcommand{\lx}{\lambda}

\def\Hc{{\cal H}}
\newcommand{\lb}{\bar{\lambda}}

\newcommand{\drho}{\delta \rho}

\newcommand{\dOmega}{{\delta \Omega}}
\newcommand{\dgamma}{{\delta \gamma}}

\newcommand{\bp}{\mathbf{p}}
\newcommand{\bq}{\mathbf{q}}
\newcommand{\bk}{\mathbf{k}}
\newcommand{\bl}{\mathbf{l}}
\newcommand{\br}{\mathbf{r}}

\DeclareSymbolFont{mathscrUC}{U}{rsfs}{m}{n}  % Formal Script for uppercase
\DeclareSymbolFont{mathscrLC}{OT1}{pzc}{m}{n} % Chancery for lowercase

\makeatletter

\DeclareRobustCommand*{\mathscr}[1]{\gdef\F@ntPrefix{mathscr@char@}%
  \@EachCharacter #1\@EndEachCharacter}
\long\def\DoLongFutureLet #1#2#3#4{% 
   \def\@FutureLetDecide{#1#2\@FutureLetToken
      \def\@FutureLetNext{#3}\else
      \def\@FutureLetNext{#4}\fi\@FutureLetNext}
   \futurelet\@FutureLetToken\@FutureLetDecide}
\def\DoFutureLet #1#2#3#4{\DoLongFutureLet{#1}{#2}{#3}{#4}}
\def\@EachCharacter{\DoFutureLet{\ifx}{\@EndEachCharacter}%
   {\@EachCharacterDone}{\@PickUpTheCharacter}}
\def\m@keCharacter#1{\csname\F@ntPrefix#1\endcsname}
\def\@PickUpTheCharacter#1{\m@keCharacter{#1}\@EachCharacter}
\def\@EachCharacterDone \@EndEachCharacter{}

\DeclareMathSymbol{\mathscr@char@A}{\mathord}{mathscrUC}{`A}
\DeclareMathSymbol{\mathscr@char@B}{\mathord}{mathscrUC}{`B}
\DeclareMathSymbol{\mathscr@char@C}{\mathord}{mathscrUC}{`C}
\DeclareMathSymbol{\mathscr@char@D}{\mathord}{mathscrUC}{`D}
\DeclareMathSymbol{\mathscr@char@E}{\mathord}{mathscrUC}{`E}
\DeclareMathSymbol{\mathscr@char@F}{\mathord}{mathscrUC}{`F}
\DeclareMathSymbol{\mathscr@char@G}{\mathord}{mathscrUC}{`G}
\DeclareMathSymbol{\mathscr@char@H}{\mathord}{mathscrUC}{`H}
\DeclareMathSymbol{\mathscr@char@I}{\mathord}{mathscrUC}{`I}
\DeclareMathSymbol{\mathscr@char@J}{\mathord}{mathscrUC}{`J}
\DeclareMathSymbol{\mathscr@char@K}{\mathord}{mathscrUC}{`K}
\DeclareMathSymbol{\mathscr@char@L}{\mathord}{mathscrUC}{`L}
\DeclareMathSymbol{\mathscr@char@M}{\mathord}{mathscrUC}{`M}
\DeclareMathSymbol{\mathscr@char@N}{\mathord}{mathscrUC}{`N}
\DeclareMathSymbol{\mathscr@char@O}{\mathord}{mathscrUC}{`O}
\DeclareMathSymbol{\mathscr@char@P}{\mathord}{mathscrUC}{`P}
\DeclareMathSymbol{\mathscr@char@Q}{\mathord}{mathscrUC}{`Q}
\DeclareMathSymbol{\mathscr@char@R}{\mathord}{mathscrUC}{`R}
\DeclareMathSymbol{\mathscr@char@S}{\mathord}{mathscrUC}{`S}
\DeclareMathSymbol{\mathscr@char@T}{\mathord}{mathscrUC}{`T}
\DeclareMathSymbol{\mathscr@char@U}{\mathord}{mathscrUC}{`U}
\DeclareMathSymbol{\mathscr@char@V}{\mathord}{mathscrUC}{`V}
\DeclareMathSymbol{\mathscr@char@W}{\mathord}{mathscrUC}{`W}
\DeclareMathSymbol{\mathscr@char@X}{\mathord}{mathscrUC}{`X}
\DeclareMathSymbol{\mathscr@char@Y}{\mathord}{mathscrUC}{`Y}
\DeclareMathSymbol{\mathscr@char@Z}{\mathord}{mathscrUC}{`Z}
\DeclareMathSymbol{\mathscr@char@a}{\mathord}{mathscrLC}{`a}
\DeclareMathSymbol{\mathscr@char@b}{\mathord}{mathscrLC}{`b}
\DeclareMathSymbol{\mathscr@char@c}{\mathord}{mathscrLC}{`c}
\DeclareMathSymbol{\mathscr@char@d}{\mathord}{mathscrLC}{`d}
\DeclareMathSymbol{\mathscr@char@e}{\mathord}{mathscrLC}{`e}
\DeclareMathSymbol{\mathscr@char@f}{\mathord}{mathscrLC}{`f}
\DeclareMathSymbol{\mathscr@char@g}{\mathord}{mathscrLC}{`g}
\DeclareMathSymbol{\mathscr@char@h}{\mathord}{mathscrLC}{`h}
\DeclareMathSymbol{\mathscr@char@i}{\mathord}{mathscrLC}{`i}
\DeclareMathSymbol{\mathscr@char@j}{\mathord}{mathscrLC}{`j}
\DeclareMathSymbol{\mathscr@char@k}{\mathord}{mathscrLC}{`k}
\DeclareMathSymbol{\mathscr@char@l}{\mathord}{mathscrLC}{`l}
\DeclareMathSymbol{\mathscr@char@m}{\mathord}{mathscrLC}{`m}
\DeclareMathSymbol{\mathscr@char@n}{\mathord}{mathscrLC}{`n}
\DeclareMathSymbol{\mathscr@char@o}{\mathord}{mathscrLC}{`o}
\DeclareMathSymbol{\mathscr@char@p}{\mathord}{mathscrLC}{`p}
\DeclareMathSymbol{\mathscr@char@q}{\mathord}{mathscrLC}{`q}
\DeclareMathSymbol{\mathscr@char@r}{\mathord}{mathscrLC}{`r}
\DeclareMathSymbol{\mathscr@char@s}{\mathord}{mathscrLC}{`s}
\DeclareMathSymbol{\mathscr@char@t}{\mathord}{mathscrLC}{`t}
\DeclareMathSymbol{\mathscr@char@u}{\mathord}{mathscrLC}{`u}
\DeclareMathSymbol{\mathscr@char@v}{\mathord}{mathscrLC}{`v}
\DeclareMathSymbol{\mathscr@char@w}{\mathord}{mathscrLC}{`w}
\DeclareMathSymbol{\mathscr@char@x}{\mathord}{mathscrLC}{`x}
\DeclareMathSymbol{\mathscr@char@y}{\mathord}{mathscrLC}{`y}
\DeclareMathSymbol{\mathscr@char@z}{\mathord}{mathscrLC}{`z}

\makeatother

\relax

\title{The dark matter bispectrum from effective viscosity and one-particle irreducible vertices}

\author[a]{Stefan Floerchinger,}
\author[b]{Mathias Garny,}
\author[c]{Aris Katsis,}
\author[c]{Nikolaos Tetradis,}
\author[d]{Urs Achim Wiedemann}

\affiliation[a]{Institut f\"{u}r Theoretische Physik, Universit\"{a}t Heidelberg, Philosophenweg 16, 69120 Heidelberg, Germany}
\affiliation[b]{Physik Department T31, Technische Universit\"{a}t M\"{u}nchen, James-Franck-Stra\ss e 1, D-85748 Garching, Germany}
\affiliation[c]{Department of Physics, University of Athens, University Campus, Zographou 157 84, Greece}
\affiliation[d]{Theoretical Physics Department, CERN,  CH-1211 Gen\`eve 23, Switzerland}

\emailAdd{stefan.floerchinger@thphys.uni-heidelberg.de}
\emailAdd{mathias.garny@tum.de}
\emailAdd{ariskatsis@phys.uoa.gr}
\emailAdd{ntetrad@phys.uoa.gr}
\emailAdd{urs.wiedemann@cern.ch}

\abstract{
Dark matter evolution during the process of
cosmological structure formation can be described in 
terms of a one-particle irreducible effective action at a characteristic 
scale $k_m$ and a loop expansion below this scale, 
based on the effective propagators and vertices. 
We calculate the form of the effective vertices and compute 
the bispectrum of density perturbations within a one-loop approximation. 
We find that the effective vertices play a subdominant role as compared to 
the effective viscosity and sound velocity that modify 
the (inverse) propagators. 
For the bispectrum we reproduce the results of 
standard perturbation theory in the range where
it is applicable, and find a slightly improved agreement with $N$-body simulations
at larger wavenumbers.}

\begin{document}

\begin{flushright}
{\small
CERN-TH-2019-122\\
TUM-HEP-1215/19\\
July 24, 2019\\
}
\end{flushright}

\maketitle

%%%%%%%%%%%%%%%%%%%%%%%%%%%%%%%%%%%%%%%%%%%%%%%%%
\section{Introduction} \label{intro}
%%%%%%%%%%%%%%%%%%%%%%%%%%%%%%%%%%%%%%%%%%%%%%%%%

The distribution of dark matter in our Universe arises dynamically from the evolution of initially small perturbations under the influence of gravity. Its precise  
description is of great interest as it can provide constraints on cosmological models~\cite{Amendola:2012ys,Abell:2009aa,Levi:2013gra,Dawson:2015wdb, Ata:2017dya}. 
It is a generic feature of cosmological structure formation that initially small deviations from average background fields grow non-perturbatively large at late time and
small length scales. On these galactic scales, calculations have to rely entirely on non-perturbative techniques such as CPU-intensive $N$-body 
simulations ~\cite{Kuhlen:2012ft, Vogelsberger:2014dza, Schneider:2015yka}. However, on sufficiently large length scales, as those of the baryon acoustic oscillations (BAO),
where density contrasts do not exceed order unity, cosmological perturbation theory has demonstrated its ability to complement $N$-body simulations~\cite{Scoccimarro:1995if, bernardeau0, bernardeau, Carlson:2009it, Audren:2011ne, CrSc1, Taruya:2012ut,CrSc2, Bernardeau:2011vy, nonlinear,Crocce:2007dt, Smith:2007gi, Sherwin:2012nh, Senatore:2014via, Blas:2015qsi, Blas:2016sfa}.
Beyond providing conceptual insight into how cosmological structures grow, such analytic approaches are of use for fast scans of classes of cosmological models and initial conditions, where 
$N$-body simulations face numerical limitations.  

In general, cosmological perturbation theory starts from the collisionless Vlasov-Boltzmann equation which can be  rewritten as a hierarchy of equations for the
moments or cumulants of the dark matter phase-space distribution~\cite{bernardeau0}. Truncating this hierarchy at the lowest moments results in equations of motion
governing the dynamics of an ideal and pressureless fluid. Taking into account the next moment of the distribution function introduces 
a shear tensor. In principle, even at large length scales, the precise calculation of the dark matter distribution from any such truncation of moment equations is 
complicated by the fact that nonlinear terms in the evolution couple short and long wavelengths. In practice, for the single stream approximation this becomes an obstacle at a characteristic momentum scale 
$k_{nl}\simeq 0.3\, h/$Mpc when nonlinearities start dominating the evolution. The onset of this breakdown of a perturbative description is seen for instance in the 
increased UV sensitivity of the higher-order corrections to the propagation of long-wavelength 
fluctuations, see e.g.~\cite{Blas1}. 

Different approaches have been suggested for an improved treatment of the UV sensitivity of cosmological perturbation theory. In particular,  the truncated set of moment equations 
may be regarded as defining an effective description, applicable below some momentum scale $k_m$. In the so-called effective field theory approach
to large scale structure, UV physics above the scale $k_m$ is parametrized in terms of additional effective couplings that are not predictable within the effective theory, but 
are fixed by comparing the calculated correlation
functions with either $N$-body simulations or observations \cite{Baumann:2010tm, Carrasco:2012cv, Porto:2013qua, Baldauf:2014qfa, Angulo:2014tfa, Foreman:2015lca, Baldauf:2015aha, Assassi:2015jqa, Abolhasani:2015mra,Fuhrer:2015cia,Chudaykin:2019ock}. These additional couplings act like `counterterms' in that they cancel the UV part of the perturbative loop corrections to various observables.

A somewhat more ambitious approach could consist in understanding also the scale dependence of these additional couplings by formulating the renormalization group (RG) which they obey. 
To pursue such an RG program, an explicit field theoretic formulation of the problem is beneficial. Methods to write stochastic differential equations as a field theory in terms of path integrals 
have been developed in many fields of physics and are often referred to as the  Martin-Siggia-Rose formalism~\cite{MSR73}. For cosmological perturbations, a closely related field theory formulation of the 
stochastic evolution equations was first given by Matarrese and Pietroni~\cite{Max1} in terms of a one-particle irreducible effective action that depends on the relevant fields (such 
as density contrast and velocity divergence) combined into $\phi$, and on the corresponding response fields $\chi$. This effective action $\Gamma[\phi,\chi]$ was supplemented with an explicit
regulator $k_m$ in Refs.~\cite{viscousdm,Floerchinger:2016hja}, indicating that, in the spirit of the Wilsonian approach to the RG (see \cite{Berges:2000ew} for a review), 
the couplings entering $\Gamma_{k_m}[\phi,\chi]$ are obtained by integrating out the small-scale cosmological perturbations~\cite{Pietroni:2011iz,Manzotti:2014loa}. In Ref.~~\cite{Floerchinger:2016hja},
we derived the exact functional RG flow for $\Gamma_{k_m}[\phi,\chi]$, and we applied this program within a set of physically motivated approximations that we recall now. 

In principle, many effective linear and nonlinear couplings could enter $\Gamma_{k_m}[\phi,\chi]$. In practice, to arrive at manageable calculations, restricting the formulation by a guiding principle
to a few `most relevant' couplings is beneficial. In Ref.~\cite{viscousdm,Floerchinger:2016hja}, our guiding principle was the use of fluid dynamics as an effective theory. Namely, when a fluid dynamic
description is limited to the long wavelength modes (larger than $ 1/k_m$) of a system, the transport properties (such as viscosity or sound velocity) governing its evolution are not given by the state-independent
properties of the matter under consideration. Rather, effective viscosity and sound velocity in a fluid dynamic description of modes with wavenumber $k<k_m$ arise from the coupling of these 
modes to the state-dependent spectra of fluctuations at wavenumbers larger than $k_m$. This had motivated an ansatz for the 
effective action~\cite{Floerchinger:2016hja} that is limited to effective scale-dependent fluid dynamic transport properties: 
\begin{equation}
\begin{split}
\Gamma_k[\phi,\chi] = \int d\eta \Bigg[ 
&\int d^3q \,\chi_a(-\bq,\eta)\left(\delta_{ab} 
\partial_\eta+ \Omega_{ab,k}(\bq, \eta)\right) \phi_b( {\bf q},\eta) \\
&-  \int d^3r\, d^3 p \, d^3 q \, \delta^{(3)}(\br-\bp-\bq)
\gamma_{abc}(\br, \bp,\bq)
\chi_a(-\br,\eta)\phi_b(\bp,\eta)
\phi_c( \bq ,\eta) \\
& -\frac{i}{2} \int d^3q\, \chi_a(\bq,\eta) H_{ab,k}(\bq,\eta,\eta^\prime) \chi_b(\bq,\eta^\prime) +\ldots \Bigg]\, .
\end{split} 
\label{eq:truncationGamma}
\end{equation}
In Ref.~\cite{Floerchinger:2016hja}, the matrix $ \Omega_{ab,k}(\bq, \eta)$ was assumed to contain scale-dependent 
contributions from the effective viscosity and sound velocity, while the vertices $\gamma_{abc}(\br, \bp,\bq)$ were assumed for simplicity to be of the tree-level 
form, i.e. unmodified by the RG running.  If calculated by matching the tree-level propagator (or power spectrum) of the effective theory with 
the one-loop propagator (or power spectrum) for an ideal fluid, the scale dependence of  $ \Omega_{ab,k}(\bq, \eta)$ arises from integrating the
power spectrum for $k> k_m$, and --- for realistic spectra --- the dominant contribution comes from the region near $k_m$, where perturbation theory is still
applicable. This is central to the argument that not only the dynamics of sufficiently long wavelength perturbations, but also the scale dependence of the couplings
that govern this dynamics should be accessible through perturbation theory. 

If the RG running of the effective viscosity and sound velocity is accounted for, then eq.~(\ref{eq:truncationGamma}) shows very good agreement with the results of
$N$-body simulations for scales $k \lesssim 0.2\, h/{\rm Mpc}$~\cite{Floerchinger:2016hja}.  The question arises to what extent this finding would remain unchanged if 
other effective couplings were included. The possible scale dependence of the effective three-point vertices  in (\ref{eq:truncationGamma}) is of particular interest
in this context, as it relates to the measurable bispectrum. As a first step towards addressing this question, we compute here
the effective terms present in $\Gamma_{k_m}[\phi,\chi]$ through the one-loop corrections to the propagator and three-point couplings. 

The paper is organized as follows: after introducing the formalism in section~\ref{sec2}, we compute in section \ref{vertices}  the one-loop correction to the
vertices of the ideal-fluid theory and use it in order to define
the vertices of the effective theory. In section \ref{recurrenc} we 
derive recurrence relations for the kernels of the effective theory 
and solve them approximately. In section \ref{sec:num} we compute the
bispectrum numerically. First we check that we can reproduce the results of 
standard perturbation theory (SPT) and then carry out an improved
calculation of the bispectrum using our effective theory.
We also compare our results with N-body simulations.
In section \ref{conclusions} we present our conclusions.

\section{The effective propagator} \label{sec2} 

Within the single stream approximation, assuming irrotational flows and neglecting the shear tensor, 
the fluid dynamic equations are written in terms of two scalar fields: the density 
perturbation $\delta\equiv\drho/\rho_m$ and the velocity divergence $\theta\equiv\vec{\nabla}\vec{v}$. Their Fourier modes are
usually included in the doublet
\begin{equation}
 \left(
\begin{array}{c}
\phi_{1}(\bk,\eta)\\ \\ \phi_{2}(\bk,\eta)
\end{array}
\right)
\equiv\left(
\begin{array}{c}
\delta_\bk(\tau)\\ \\-\dfrac{\theta_{\bk}( \tau)}{\mathcal{H}}
\end{array}
\right)\, .
\label{doublet}
\end{equation}
The dynamic equations take the form
\be
\partial_\eta \phi_a (\bk,\eta)= -\Omega_{ab}(\bk,\eta) \phi_b (\bk,\eta)+
 \int d^3 p \, d^3 q \, \delta^{(3)}(\bk-\bp-\bq)
\Gamma_{abc}(\bp,\bq,\eta)\, \phi_b (\bp,\eta)\, \phi_c(\bq,\eta)\, .
\label{eom}
\ee
Here, the conformal Hubble parameter $\Hc=\dot{a}/a$ is defined as usual in terms of the scale factor $a(\tau)$, $\Omega(\bk,\eta)=\Omega^0(\eta)$ with 
\be
\Omega_{ab}^0(\eta)=\left(
\begin{array}{cc}
~~~0 &~~~-1
\\
-\frac{3}{2} \Omega_m &~~~ 1 +\frac{\Hc'}{\Hc}
\end{array} \right)\, ,
\label{ome} \ee
and 
the prime denotes a derivative with respect to the `time' $\eta=\ln a(\tau)$. 
The nonzero elements of the interaction terms at tree-level 
$\Gamma_{abc}(\bp,\bq,\eta)=\gamma_{abc}(\bp,\bq)$ are 
\bea
\gamma_{112}(\bp,\bq)=\gamma_{121}(\bq,\bp)&=&\frac{(\bp+\bq)\bq}{2q^2} \label{gamma1}, \\
\gamma_{222}(\bp,\bq)&=&\dfrac{(\bp+\bq)^{2} \bp\cdot\bq}{2 p^{2} q^{2}}\, .\label{gamma3} 
\eea 

The evolution is particularly simple in an Einstein-de Sitter (EdS) Universe with 
$\Omega_m=1$ and $\Hc'/\Hc=-1/2$, and we concentrate on this case. 
Any $\Lambda$CDM cosmology can be mapped,
through an appropriate change of variables, to one with $\Omega_m=1$ to a very
good approximation \cite{bernardeau0}.
The retarded linear propagator $G^\text{R}_{ab}(\bk,\eta,\eta')$ satisfies
\be
\left( \delta_{ac} \, \partial_\eta + \Omega_{ac}(\bk,\eta) \right) G_{cb}^\text{R} (\bk, \eta,\eta^\prime) = \delta_{ab} \delta\left(\eta-\eta^\prime\right)\,. 
\label{linpropa}
\ee
For an EdS Universe,
 it is given by $G^\text{R}(\bk,\eta,\eta')=g^\text{R}(\eta-\eta')$
 with 
\be
g_{ab}^\text{R}(\eta-\eta^\prime) = \left[ 
\frac{e^{\eta-\eta^\prime}}{5} \begin{pmatrix} 3 && 2 \\ 3 && 2 \end{pmatrix} 
 - \frac{e^{-3(\eta - \eta^\prime) / 2}}{5} \begin{pmatrix} -2 && ~~2 \\ ~~3 && -3 \end{pmatrix}\right]  \Theta\left(\eta-\eta^\prime\right)\, ,
\label{eq2.9}
\ee
where the growing and decaying modes are visible.

The effective propagator of the low-energy theory receives loop corrections arising through mode coupling to the UV sector 
that is integrated out. These corrections can be taken into account to
a very good approximation by the ansatz~\cite{viscousdm,Floerchinger:2016hja}
\be
\Omega_{ab}(\bk,\eta)=
\Omega^0+\dOmega(\bk,\eta)=
\left(
\begin{array}{cc}
~~~0 &~~~-1
\\
-3/2 &~~~ 1/2
\end{array} \right)
+\left(
\begin{array}{cc}
~~~0 &~~~0
\\
\lambda_s \exp(2\eta) k^2&~~~ \lambda_\nu \exp(2\eta) k^2
\end{array} \right) 
\label{omeren} \ee
in terms of an effective viscosity $\nu$ and sound velocity $c^2_s$,
\be
\nu\equiv \frac{3}{4}\lambda_\nu \Hc \exp(2\eta)\, ,~~~~~~~~~~~~~~~
c^2_s \equiv \lambda_s \Hc^2 \exp(2\eta)\, .
\label{anu} 
\ee
Both quantities become relevant only in the recent past, as can be seen from the very strong time dependence.

The form of $\Omega_{ab}(\bk,\eta)$ indicates that the solution of the
linearized evolution (\ref{eom}) satisfies $\phi_2(\bk,\eta)=\partial_\eta \phi_1(\bk,\eta)$.
For $\phi_1(\bk,\eta)=\delta(\bk,\eta)$, this leads to the differential equation 
\be
\partial^2_\eta \delta(\bk,\eta)
+\left(\frac{1}{2}+\lambda_\nu k^2 \exp(2\eta) \right)\partial_\eta \delta(\bk,\eta)
+\left(-\frac{3}{2}+\lambda_s k^2 \exp(2\eta) \right) \delta(\bk,\eta)=0\, ,
\label{eqphi1} \ee
whose exact solution in terms of growing and decaying modes 
\be
\delta(\bk,\eta)=c_1\,f_g(\bk,\eta)+c_2\,f_d(\bk,\eta)
\label{lindelta} \ee
can be expressed in terms of hypergeometric functions 
$_1F_1(a,b,z)$,
\bea
f_g(\bk,\eta)&=& \exp(\eta)\,
_1 F_1\left(\frac{1}{2}+ \frac{\lambda_s}{2\lambda_\nu}, \frac{9}{4}, -\frac{\lambda_\nu}{2} k^2  \exp(2 \eta) \right)\, ,
\label{solfg} \\
f_d(\bk,\eta)&=& \exp(-3\eta/2)\,
_1 F_1 \left(-\frac{3}{4}+ \frac{ \lambda_s}{2 \lambda_\nu},  -\frac{1}{4}, -\frac{\lambda_\nu}{2}  k^2 \exp(2 \eta) \right)\, .
\label{solfd} \eea
The full retarded propagator takes then the form
\bea
G_{ab}^\text{R}(\bk,\eta,\eta^\prime)&=& \frac{\Theta\left(\eta-\eta^\prime\right)}{f_g(\eta')\,f_d'(\eta')-f_g'(\eta')\,f_d(\eta')}
\nonumber \\
&& \times \left[  \begin{pmatrix} f_d'(\eta')\,f_g(\eta) &&- f_d(\eta')\,f_g(\eta) \\  f_d'(\eta')\,f_g'(\eta) &&- f_d(\eta')\,f_g'(\eta) \end{pmatrix} 
-  \begin{pmatrix} f_g'(\eta')\,f_d(\eta) &&- f_g(\eta')\,f_d(\eta) \\  f_g'(\eta')\,f_d'(\eta) &&- f_g(\eta')\,f_d'(\eta) \end{pmatrix}   \right]
\, .
\label{viscousprop}
\eea
Here, $f_{d,g}^\prime\equiv \partial_\eta f_{d,g}$, and the $\bk$-dependence of $f_g$ and $f_d$ is not made explicit for notational reasons. 
In contrast to the bare propagator in (\ref{eq2.9}), $G_{ab}^\text{R}(\bk,\eta,\eta^\prime)$ depends separately on 
$\eta$ and $\eta'$ because of the $\eta$-dependence of $\dOmega$.

\begin{figure}{t}
\centering
\includegraphics[width=0.6\textwidth]{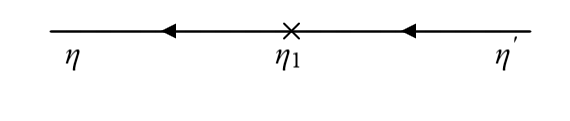}
\includegraphics[width=0.6\textwidth]{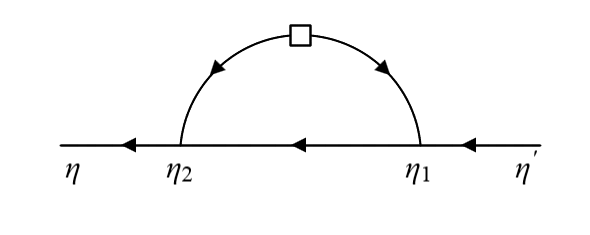}
\caption{The effective contribution to the propagator.
}
\label{fig1}
\end{figure}

The propagator can be expanded in powers of $k^2$, with the first two terms given by 
\be
G_{ab}^\text{R}(\bk,\eta,\eta^\prime)=g_{ab}^\text{R}(\eta-\eta^\prime)- (\lambda_\nu+\lambda_s) \exp(2\eta)\, k^2 \, 
\frac{1}{45} \begin{pmatrix} 3 && 2 \\ 9 && 6 \end{pmatrix}  \exp(\eta-\eta'),
\label{propexp} \ee
where we have kept the leading contribution for large $\eta-\eta'$. As expected, the correction to the `tree-level' propagator 
comes entirely from the growing mode. The same expression can be obtained if one considers the correction to the 
propagator arising from a `mass insertion' $-\dOmega(\bk,\eta_1)$, with the external legs corresponding to tree-level propagators
of the form (\ref{eq2.9}), as shown in the upper diagram of fig. \ref{fig1}. 
The integration over the internal `time' $\eta_1$ reproduces eq. (\ref{propexp}) for a large difference $\eta-\eta'$  between initial and final times.
Again, the result arises entirely from the growing mode. 

As we have discussed above, we view the effective couplings as arising from the UV modes that have been integrated out.
For this correspondence to be meaningful, the correction to the tree-level propagator in eq. (\ref{propexp}) must have the same form as the 
leading one-loop correction to the propagator in the ideal fluid theory, depicted in the lower diagram of fig. \ref{fig1}, which scales $\sim k^2$.
Two conditions must be imposed: the loop integral must be evaluated 
with a lower cutoff equal to $k_m$, and the calculation must be performed for the  growing mode. The 
calculation is similar to that of \cite{CrSc2}, apart from the presence of the cutoff. 
The result, already quoted in \cite{viscousdm}, is 
\be
G^\text{R}_{ab}(\bk, \eta,\eta^\prime)
= g^\text{R}_{ab}(\eta-\eta^\prime) - \sigma_{d}^2(\eta) k^2
\begin{pmatrix} \frac{61}{350} && \frac{61}{525} \\ \frac{27}{50} && \frac{9}{25} \end{pmatrix}
 \exp(\eta-\eta^\prime)\, ,
\label{GG}
\ee
where
\be
\sigma^2_d(\eta)  \equiv  \frac{4\pi}{3} \int_{k_m}^\infty dq\, P^L(q,\eta)\, ,
\label{smallk} \ee
with $P^L(q,\eta)$ the linear power spectrum at time $\eta$. 
For an Einstein-de Sitter Universe, the time-dependence of eqs. (\ref{propexp}) and (\ref{GG}) can be matched. Moreover, the form of the two
matrices can also be matched with a few percent accuracy if we set $\lambda_\nu+\lambda_s=2.7 \sigma^2_d$,  with $\sigma^2_d$ the present
value $(\eta=0)$ of the parameter defined in eq. (\ref{smallk}). Values for $\lambda_\nu$ and $\lambda_s$ could be fixed separately 
by considering also the decaying mode~\cite{viscousdm,Floerchinger:2016hja}, but for the purpose of the present work, we constrain only the sum 
$\lambda_\nu+\lambda_s$ from the growing mode in (\ref{GG}). 

\section{The effective vertices} \label{vertices}

\begin{figure}[t]
\centering
$$
\includegraphics[width=0.42\textwidth]{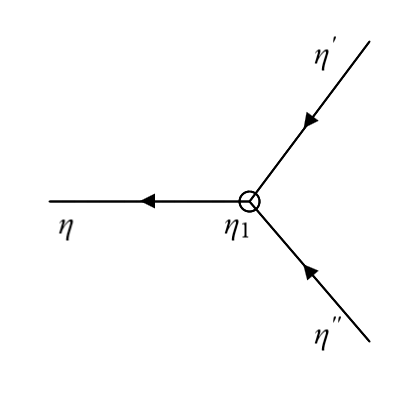}\hspace*{0.02\textwidth}
\includegraphics[width=0.42\textwidth]{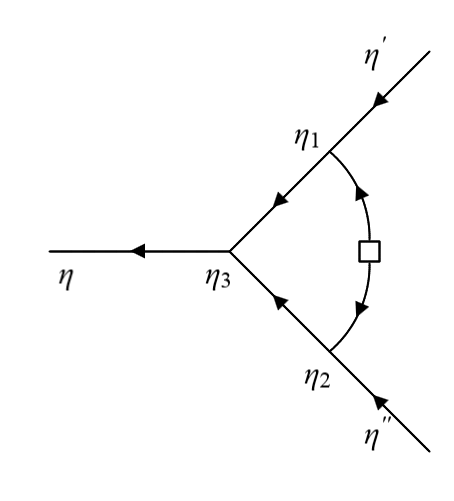}\hspace*{0.02\textwidth}
$$
\caption{The effective contribution to the three-point vertices.
}
\label{fig2}
\end{figure}

In analogy to what was done for the propagator in the previous
section, we would like to define effective three-point couplings that
generalize the tree-level ones (\ref{gamma1}), (\ref{gamma3}) and account for 
the effect of mode coupling to the UV sector. 
A typical contribution to such couplings is depicted in the
second diagram of fig. \ref{fig2}, where the momentum integration 
is performed with a lower cutoff $k_m$. The external legs of the 
diagram are not amputated, but correspond to tree-level propagators.
We assume that the incoming propagators correspond to the growing mode,
which allows us to take the limit $\eta',\eta''\to -\infty$. On the 
other hand, all other propagators include the decaying mode. 
There are two more diagrams, not depicted in fig. \ref{fig2}. 
The second loop diagram has a similar structure,
but the initial power spectrum (depicted by the small square) 
is inserted in the internal
line connecting $\eta_1$ and $\eta_3$. In the third diagram 
the square is inserted in the line connecting $\eta_2$ and $\eta_3$. 
Adding the three contributions and performing 
the integration
over the internal times $\eta_1$, $\eta_2$, $\eta_3$ leads to an expression
that depends only on $\eta$ and the external momenta. 
This expression 
can be mapped onto the one resulting from the first diagram of fig. \ref{fig2},
where the open circle denotes an effective vertex 
\be
\Gamma_{abc}(\bp,\bq,\eta)=\gamma_{abc}(\bp,\bq)+\dgamma_{abc}(\bp,\bq,\eta).
\label{vertexdef} \ee

The procedure outlined above results in effective vertices that are local
in time. In general, the effective three-point couplings include 
contributions nonlocal in time, so that a recipe is required in order
to project them onto a local expression. In \cite{Floerchinger:2016hja} this
was achieved for the propagator through an appropriate Laplace transform.
Here we perform the projection by considering only growing-mode incoming
propagators. 
Formally, this procedure has an element of arbitrariness, but it will be shown to be self-consistent and to capture the most important contributions to the  effective vertices.

The calculation of the effective vertices is rather technical. We 
present some details in appendix \ref{appendixa}. The result has two
characteristic properties:
\begin{enumerate}
\item
The corrections $\dgamma_{abc}$ to the tree-level vertices include a factor
of the parameter 
(\ref{smallk}), in complete analogy to what was found for the effective
propagator~\cite{viscousdm,Floerchinger:2016hja}. This introduces a time-dependence $\sim \exp(2\eta)$, and it
indicates that the influence of the UV sector becomes important only in
the recent past.
\item
The projection onto growing incoming modes, proportional to $\phi = (1,1)$, in  
the first diagram of fig. \ref{fig2} results in the linear combinations $\dgamma_{a11}+\dgamma_{a12}+\dgamma_{a21}+\dgamma_{a22}$, with $a=1,2$.
Remarkably, the second diagram of fig. \ref{fig2} leads to an expression
with only two independent elements that can be matched to the above
linear combination of effective vertices. On the one hand, this shows that the projection to growing modes leads to a closed system of equations that is
of manageable complexity since it contains only two unknown functions.
On the other hand, this means that our procedure
cannot constrain each element $\dgamma_{abc}$ individually.  
\end{enumerate}
We obtain 
\bea
(\dgamma_{111}+\dgamma_{112}+\dgamma_{121}+\dgamma_{122})(\bp,\bq,\eta)&=&
\sigma^2_d(\eta) k^2 \, R_1(\bp,\bq)\, ,
\label{dg1} \\ 
(\dgamma_{211}+\dgamma_{212}+\dgamma_{221}+\dgamma_{222})(\bp,\bq,\eta)&=&
\sigma^2_d(\eta) k^2  \, R_2(\bp,\bq)\, ,
\label{dg2} \eea
where
\bea
R_1(\bp,\bq)&=&
-\frac{5815\, \bk^6+5392\, \bk^4 (\bp^2+\bq^2) -32198\, \bk^2 (\bp^4+\bq^4)
}{123480 \, \bk^2 \bp^2 \bq^2}
\nonumber \\
&& -\frac{-25040\, \bk^2 \bp^2 \bq^2+20991 (\bp^6- \bp^4 \bq^2
- \bp^2 \bq^4+ \bq^6)}{123480 \, \bk^2 \bp^2 \bq^2}\, ,
\label{R1} \\
R_2(\bp,\bq)&=&
-\frac{74\, \bk^4+307\, \bk^2 (\bp^2+\bq^2)
+1974\,  \bp^2 \bq^2-381\, (\bp^4+\bq^4)}{1176\,  \bp^2 \bq^2}\, ,
\label{R2} \eea
with $\bp$, $\bq$ the incoming momenta and
$\bk=\bp+\bq$ the outgoing one. Both $R_1$ and $R_2$ take finite values for 
$\bp^2=\bk^2$, $\bq^2=0$, or $\bq^2=\bk^2$, $\bp^2=0$, so that 
no spurious infrared singularities appear. 
For an Einstein-de Sitter Universe, the leading time dependence of the quantity $\sigma^2_d(\eta)$, defined in eq. (\ref{smallk}), is 
$\sigma^2_d(\eta)=\sigma^2_d \exp(2\eta)$, with $\sigma^2_d$ its value today
($\eta=0$). 

\section{Recurrence relations}
\label{recurrenc}

A standard method for the solution of eq. (\ref{eom}) is to expand the 
density contrast $\delta = \phi_1$ and rescaled velocity divergence 
$-\theta/{\cal H} = \phi_2$
in powers of the Fourier modes $\delta_{\bq_n}(\eta_0)$ of the initial density perturbations at $\eta=\eta_0$~\cite{goroff,bernardeau0}:
\be
  \phi_a(\bk,\eta) = \sum_{n=1}^\infty \int d^3q_1 \cdots d^3q_n \, (2\pi)^3 \delta^{(3)}\left( \bk-\sum_i\bq_i \right) F_{n,a}(\bq_1,\dots ,\bq_n,\eta )
  \delta_{\bq_1}(\eta_0)\cdots\delta_{\bq_n}(\eta_0)\;.
\label{expkern}
\ee
Inserting into eq. (\ref{eom}) gives an evolution equation 
for the kernels $F_{n,a}$
\bea
  &&(\partial_{\eta}\delta_{ab}+\Omega_{ab}(\bk,\eta))F_{n,b}(\bq_1,\dots ,\bq_n,\eta)
  = 
  \nonumber \\
&& \sum_{m=1}^{n-1} \Gamma_{abc}(\bq_1+\cdots+\bq_m,\bq_{m+1}+\cdots+\bq_n,\eta) 
   F_{m,b}(\bq_1,\dots,\bq_m,\eta)F_{n-m,c}(\bq_{m+1},\dots,\bq_n,\eta)  , \quad \label{kernels}
\eea
where the right-hand side is understood to be symmetrized w.r.t. arbitrary permutations of the $\bq_i$, and $\bk=\sum_i \bq_i$.
When neglecting the viscosity and sound-velocity terms and the corrections 
$\delta\gamma_{abc}$ to the effective
vertices, the solution is of the form 
$F_{n,1}=\exp(n\eta)F_n$ and $F_{n,2}=\exp(n\eta)G_n$ \cite{bernardeau0}. 
Unfortunately,
the time dependence introduced through the effective terms does not allow for such an
exact factorization.The exact determination of the kernels is 
possible through the numerical 
solution of the first-order differential equation (\ref{kernels}).
However, for modes of low momentum an explicit solution is still
possible at order $k^2$. 

The kernels $F_{1,a}$ can be obtained from 
the linear evolution. For $\eta\to-\infty$, the matrix $\delta\Omega$, defined
in eq. (\ref{omeren}), vanishes and the evolution becomes the standard one,
with the known growing and decaying modes. This leads to the identification 
$F_{1,1}(\bk,\eta)=f_g(\bk,\eta)$, $F_{1,2}(\bk,\eta)=\partial_\eta f_g(\bk,\eta)$, 
with $f_g(\bk,\eta)$ given by eq. (\ref{solfg}).
For sufficiently low $k^2$, the viscosity and sound-velocity terms are subleading
during the whole linear evolution until today. The kernels can be approximated by expanding 
the hypergeometric functions, with the result
\bea
F_{1,1}(\bk,\eta)&\simeq& \exp(\eta)-\frac{1}{9}(\lx_s+\lx_\nu) k^2\exp(3\eta)\, ,
\label{approxf11} \\
F_{1,2}(\bk,\eta)&\simeq&
\exp(\eta)-\frac{1}{3}(\lx_s+\lx_\nu)k^2\exp(3\eta)\, .
\label{approxf12}
\eea

We can express the matrix $\delta \Omega$ 
of eq. (\ref{omeren})
as $\delta \Omega_{ab}=(\sigma^2_d\,k^2_m) \, (k^2/k^2_m)\exp(2\eta) \lb_{ab}$.
The product $\sigma^2_d\,k^2_m$ takes values close to 0.5 for 
$k_m$ in the range $(0.4-1)\, h/$Mpc. In the low-energy effective theory, all
modes satisfy $k^2/k^2_m \leq 1$, while $\exp(2\eta)\leq 1$, for $\eta\leq 0$.
The entries of
the matrix $\lb_{ab}$ are dimensionless constants: 
$\lb_{11}=\lb_{12}=0,\lb_{21}=\lambda_s/\sigma^2_d\equiv\lb_s,
\lb_{22}=\lambda_\nu/\sigma^2_d\equiv\lb_\nu$,
with $\lb_s+\lb_\nu\simeq 2.7$ according to the matching of expressions
(\ref{propexp}) and (\ref{GG}) for the effective propagator.
Similarly, we 
parameterize the effective vertices $\delta\gamma_{abc}$ of eq. (\ref{vertexdef})
as $\delta\gamma_{abc}=(\sigma^2_d\,k^2_m) \, (k^2/k^2_m) \exp(2\eta) g_{abc}$,
where $g_{abc}(\bp,\bq)$ are dimensionless ratios of external momenta
that satisfy 
$(g_{111}+g_{112}+g_{121}+g_{122})(\bp,\bq)=R_1(\bp,\bq)$ and 
$(g_{211}+g_{212}+g_{221}+g_{222})(\bp,\bq)=R_2(\bp,\bq)$,
consistent with eqs.~(\ref{dg1}) and (\ref{dg2}).

Within the low-energy effective theory, 
keeping terms up to order $k^2$ is a good approximation
for all modes even today. In order to find an approximate solution
for the kernels, we
define the functions $F^I_{n,a}$, $F^{II}_{n,a}$ according to 
\bea
F_{n,a}(\bq_1,\dots ,\bq_n,\eta)&\simeq& F^I_{n,a}(\bq_1,\dots ,\bq_n)\exp(n\eta) \nonumber \\
&& +F^{II}_{n,a}(\bq_1,\dots ,\bq_n)\sigma^2_d \, k^2 \exp((n+2)\eta)\, .
\label{expfg} \eea
Up to first order in $k^2$, we obtain the standard recurrence relation
\bea
  &&(n\,\delta_{ab}+\Omega^0_{ab})F^I_{n,b}(\bq_1,\dots ,\bq_n)=
  \nonumber \\
  && \sum_{m=1}^{n-1} \gamma_{abc}(\bq_1+\cdots+\bq_m,\bq_{m+1}+\cdots+\bq_n) 
  \,F^I_{m,b}(\bq_1,\dots,\bq_m)F^I_{n-m,c}(\bq_{m+1},\dots,\bq_n), \; 
\label{kernelsI}
\eea
along with the new one
\bea
  &&((n+2)\,\delta_{ab}+\Omega^0_{ab})F^{II}_{n,b}(\bq_1,\dots ,\bq_n)
+ \lb_{ab}F^{I}_{n,b}(\bq_1,\dots ,\bq_n)=
\nonumber \\
  \sum_{m=1}^{n-1}  &\Bigl[&
  \gamma_{abc}(\bq_1+\cdots+\bq_m,\bq_{m+1}+\cdots+\bq_n) 
  F^I_{m,b}(\bq_1,\dots,\bq_m) F^{II}_{n-m,c}(\bq_{m+1},\dots,\bq_n) \Bigr.
\nonumber \\
  &+& 
  \gamma_{abc}(\bq_1+\cdots+\bq_m,\bq_{m+1}+\cdots+\bq_n) 
  F^{II}_{m,b}(\bq_1,\dots,\bq_m) F^I_{n-m,c}(\bq_{m+1},\dots,\bq_n) 
  \nonumber \\
\Bigl.  &+& 
  g_{abc}(\bq_1+\cdots+\bq_m,\bq_{m+1}+\cdots+\bq_n) 
  F^I_{m,b}(\bq_1,\dots,\bq_m) F^I_{n-m,c}(\bq_{m+1},\dots,\bq_n) \Bigr].
    \; 
\label{kernelsII}
\eea
We discuss now the form of the kernels to understand the nature of the higher-order
corrections. Comparing the ansatz (\ref{expfg}) with (\ref{approxf12}), (\ref{approxf11}), we write
\bea
F^I_{1,1}&=&F^I_{1,2}=1\, ,
\label{FI1} \\
F^{II}_{1,1}&=&(-1/9)(\lb_s+\lb_\nu)\, ,
\label{FII1} \\
F^{II}_{1,2}&=&(-1/3)(\lb_s+\lb_\nu)\, ,
\label{FII2}
\eea
which is consistent with (\ref{kernelsII}). This shows that the linear evolution is affected at late times by the effective 
viscosity and sound velocity, and that the degeneracy 
between $\lb_\nu$ and $\lb_s$ is not broken at the linear level.
Use of the phenomenological values $\lb_s+\lb_\nu \simeq 2.7$, $\sigma^2_d k^2_m \simeq 0.5$ in eq. (\ref{expfg}) indicates that the velocity 
field suffers the maximal effect, of order of $50\%$, for $\eta \simeq 0$ and $k^2\simeq k^2_m$. The maximal effect on the density field is much weaker, 
of the order of $15\%$. For modes with lower $k$, the
viscosity and sound-velocity corrections
are suppressed by an additional factor $\sim k^2/k^2_m$.

For $n=2$, eq. (\ref{kernelsI}) gives the standard relation
\bea
F_{2,1}^I(\bp,\bq)&=& \frac{5\left(\gamma_{112}(\bp,\bq)+\gamma_{121}(\bp,\bq)\right)
+2\gamma_{222}(\bp,\bq)}{7}  \nonumber \\
&=&
\frac{5}{7}+\frac{(p^2+q^2)\,\bp\cdot \bq}{2p^2 q^2}
+\frac{2}{7}\frac{(\bp\cdot \bq)^2 }{p^2 q^2}\, ,
\label{FI21} \\
F_{2,2}^I(p,q)&=& \frac{3\left(\gamma_{112}(\bp,\bq)+\gamma_{121}(\bp,\bq)\right)
+4\gamma_{222}(\bp,\bq)}{7} \nonumber \\
&=&
\frac{3}{7}+\frac{(p^2+q^2)\, \bp\cdot \bq}{2p^2 q^2}
+\frac{4}{7}\frac{(\bp\cdot \bq)^2 }{p^2 q^2}\, ,
\label{FI22} \eea
while eq. (\ref{kernelsII}) gives
\bea
F_{2,1}^{II}(\bp,\bq)&=&
\frac{3}{11}\left[R_1(\bp,\bq)+ \left(\gamma_{112}(\bp,\bq)+\gamma_{121}(\bp,\bq)\right)(F_{1,1}^{II}+F_{1,2}^{II})\right]
\nonumber \\
&&+
\frac{2}{33}\left[R_2(\bp,\bq)-\lb_s 
F_{2,1}^I(\bp,\bq)-\lb_\nu F_{2,2}^I(\bp,\bq)+2\gamma_{222}(\bp,\bq)F_{1,2}^{II}\right]\, ,
\label{FII21f} \\
F_{2,2}^{II}(\bp,\bq)&=&
\frac{1}{11}\left[R_1(\bp,\bq)+ \left(\gamma_{112}(\bp,\bq)+\gamma_{121}(\bp,\bq)\right)(F_{1,1}^{II}+F_{1,2}^{II})\right]
\nonumber \\
&&+
\frac{8}{33}\left[R_2(\bp,\bq)-\lb_s F_{2,1}^I(\bp,\bq)-\lb_\nu F_{2,2}^I(\bp,\bq)+2\gamma_{222}(\bp,\bq)F_{1,2}^{II}\right]\, .
\label{FII22f} \eea
Remarkably, with the help of eqs.~(\ref{FI21}), (\ref{FI22}), the kernels $F^{II}_{2,a}(\bp,\bq)$ can be shown to depend only
on the sums of effective vertices that define $R_a(\bp,\bq)$ 
in eqs. (\ref{R1}), (\ref{R2}), 
\bea
F_{2,1}^{II}(\bp,\bq)&=&
\frac{1}{693}
\left[
189 R_1+42 R_2
-\lb_\nu \left( 
102 \gamma_{112}+102\gamma_{121}+52\gamma_{222}\right) \right. \nonumber \\
&& \left. \qquad \qquad
-\lb_s \left( 
114 \gamma_{112}+114\gamma_{121}+40\gamma_{222}\right)\right] (\bp,\bq)\, ,
\label{FII21} \\
F_{2,2}^{II}(\bp,\bq)&=&
\frac{1}{693}
\left[
63 R_1+168 R_2
-\lb_\nu \left( 
100 \gamma_{112}+100\gamma_{121}+208\gamma_{222} \right) \right. \nonumber \\
&& \left. \qquad \qquad
-\lb_s \left( 
148 \gamma_{112}+148\gamma_{121}+160\gamma_{222} \right)
\right](\bp,\bq)\, . 
\label{FII22} \eea

Before discussing in detail the kernels $F^{II}_{2,a}$,
three remarks are in order: 
\begin{enumerate}
\item 
 The matching procedure employed in section \ref{sec2} only constrains $\lb_s+\lb_\nu\simeq 2.7$. However, this degeneracy is broken at the level of the kernels 
 $F^{II}_{2,a}$ which depend separately on $\lb_\nu$ and $\lb_s$.
For the numerical results presented in the following, we assumes that both values are of order unity. 
The dependence on the precise choice turns out to be mild (see Sec.\,\ref{sec:num}).

\item
Both $F^{I}_{n,a}(\bq_1,...,\bq_n)$ and $F^{II}_{n,a}(\bq_1,...,\bq_n)$ scale proportional
to $k^2$ in a limiting case for which the sum $\bk=\sum_i \bq_i$ of their arguments approaches zero, while the
magnitudes of the $\bq_i$ are held fixed, in accordance with the expectation due to overall momentum conservation \cite{bernardeau0}.
For the standard kernel, which is a rational function of its arguments, this property implies
that $F^{I}_{n,a}(\bq_1,...,\bq_n)\to 0$ also in the limiting case in which $k$ is held fixed while the $\bq_i$ become large.
However, this is not the case for $F^{II}_{n,a}(\bq_1,...,\bq_n)$.
The factor $\sigma^2_d \, k^2$ in (\ref{expfg}) implies that 
$F^{II}_{n,a}(\bq_1,...,\bq_n)$ approach a constant value for large $\bq_i/k$ when $k$ is held fixed.

\item  The kernels $F^{II}_{2,a}$ in eqs. (\ref{FII21}) and (\ref{FII22}) have terms proportional to $R_1$ and $R_2$ originating from the corrections 
to the effective vertices, and terms proportional to $\lb_s$, $\lb_\nu$ and $\gamma_{abc}$ that arise through the effective propagator and standard vertices. 
Direct comparisons of the functions $\gamma_{abc}(\bp,\bq)$ and $R_{a}(\bp,\bq)$ reveals that the latter are dominant in the UV, while the former
dominate for $\bp$, $\bq$ with magnitudes in the vicinity of $k$. As $R_{a}(\bp,\bq)$ absorb higher-order corrections of standard perturpation theory, this 
illustrates how the increased UV sensitivity of STP is tamed in the present formulation: The cutoff $k_m$ in the low-energy effective theory eliminates the 
UV region of momentum integrations in which effective terms, such as $R_a$, would dominate 
over the tree-level vertices. The value of the  tree level vertices $\gamma_{abc}$ sets the relative weight of $R_a$ and $\lb_s$, $\lb_\nu$ in the kernels 
(\ref{FII21}), (\ref{FII22}).
\end{enumerate}

\begin{figure}
\centering
\includegraphics[width=0.8\textwidth]{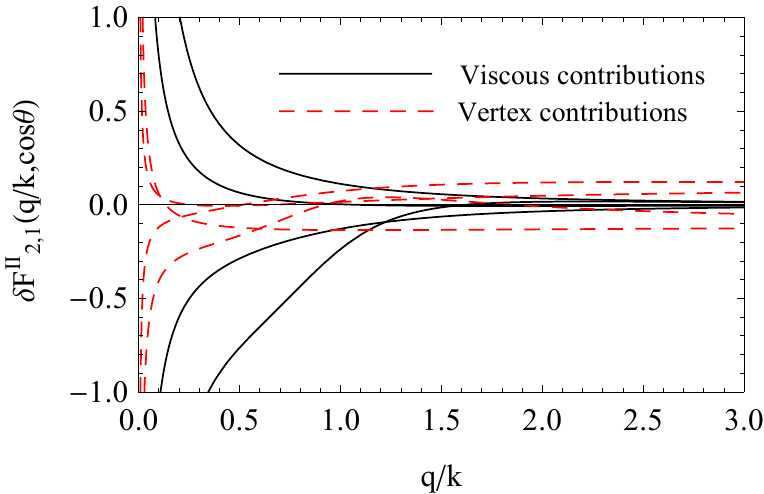}
\caption{
Contributions to the kernel $F^{II}_{2,1}(\bp,\bq)$ as given by eq.\ (\ref{FII21}) , for $\lb_s=\lb_\nu=1.4$ and for various values of 
the angle $\theta$ 
between $\bp$ and $\bq$ with $\bk=\bp+\bq$. 
The four lines in each set (solid and dashed) correspond to 
$\cos\theta$=0.8, 0.3, -0.3, -0.8.
The solid lines correspond to the contribution from the terms proportional to $\lb_s$ and $\lb_\nu$ in eq. (\ref{FII21}), while the dashed lines to the contribution from the terms proportional to $R_1$ and $R_2$. It is apparent that the first contribution dominates for $q/k\lesssim 1$, while the second one dominates for $q/k\gtrsim 1$, where it is roughly constant.
}
\label{fig3}
\end{figure}

The features we discussed above are apparent in fig. \ref{fig3}, in which
we depict contributions to the kernel $F^{II}_{2,1}(\bp,\bq)$
of eq. (\ref{FII21}) in terms of the magnitude of $\bq$ and the angle $\theta$
between $\bq$ and $\bk=\bp+\bq$. 
One sees that the contribution from terms proportional to $\lb_s$ and $\lb_\nu$ dominates
for $q/k\lesssim 1$, while the the contribution from terms proportional to $R_1$ and $R_2$ dominates for 
$q/k\gtrsim 1$, where it is roughly
constant. We are interested in the nonlinear corrections to the spectrum and 
bispectrum in the range $0.05 \lesssim k \lesssim 0.3\, h/$Mpc.
When the kernels are used for their calculation through perturbation theory, 
the momentum integrations include factors of the linear spectrum 
$P^L(q)$, which peaks at momenta below the above range. This favors 
the region of integration $q/k\lesssim 1$, in which the effective 
viscosity and sound-velocity terms dominate. The vertex corrections 
would become important for very large $q$, for which $F^{II}_{2,1}$ would
be constant and a possible UV divergence could result from an integrand 
such as $q^2\,P^L(q)$. However, this momentum 
regime is eliminated by an upper cutoff 
equal to $k_m$ in all computations within the effective theory.

It is also important to notice that, for $\lb_s$, $\lb_\nu$ of order 1, 
the magnitudes of the two types of contributions in eq. (\ref{FII21}) are set
by the tree-level vertices $\gamma_{abc}$ and their corrections $R_a$.  
The consistency of our scheme requires the vertex corrections to be subleading.
It seems reasonable then to expect that 
the effect of the vertex corrections
on all higher-order kernels is subleading to that of viscosity and 
sound velocity. We shall find support for this conclusion 
through a numerical analysis
at the level of the bispectrum in the following section. 
As a result, the calculation of higher-order kernels 
within the effective theory of our scheme can be performed by keeping the full 
matrix $\Omega_{ab}$ of eq. (\ref{omeren}), but
neglecting the vertex corrections $\delta\gamma_{abc}$ in eq. (\ref{vertexdef}).

\section{Numerical analysis}\label{sec:num}

The bispectrum within the effective theory can be computed using the expressions
familiar from standard perturbation theory (SPT), with two changes: 
\begin{enumerate} 
\item
The kernels $F_{n,a}$ are not the usual SPT kernels,
but are obtained instead by solving numerically the evolution equation \eqref{kernels}, taking the effective viscosity 
and sound velocity parameterized by $\lambda_\nu$ and $\lambda_s$, as
well as the modified vertices $\delta\gamma_{abc}$, into account. These effective
parameters depend on the scale $k_m$. 
\item 
Within the effective theory, only wavenumbers below $k_m$ contribute to
the one-loop expressions for the bispectrum. 
\end{enumerate}

The physical power and bi-spectra do not depend on the renormalization scale $k_m$. On the other hand,
the approximations leading to the effective description discussed above will in general lead to a residual 
dependence of the theoretical prediction on $k_m$. This dependence is expected to be smaller
the better the approximation captures the true power and bi-spectra. It 
can, therefore, serve as
a measure of the theoretical uncertainty, which we will quantify below. The 
setup is similar in spirit to the
dependence of perturbative predictions on the renormalization 
scale $\mu$ in quantum field theory.

An additional expected feature is that both the tree-level and one-loop
contributions, when computed within the effective theory, 
feature a sizeable dependence on $k_m$, which
approximately cancels when summing them. 
This serves as a further consistency check and validation, as
we will discuss in the following. 

For completeness, we quote the explicit expressions for the bispectrum of the density contrast up to one loop,
\bea
  B_{\rm tree}(k_1,k_2,k_3,\eta) &=& 2F_2(\bk_1,\bk_2,\eta)F_1(k_1,\eta)F_1(k_2,\eta)P_0(k_1)P_0(k_2) + 2\ {\rm permutations}\nn\\
  B_{\rm one-loop}(k_1,k_2,k_3,\eta) &=& (B_{222}+B_{321}^I+B_{321}^{II}+B_{411})(k_1,k_2,k_3,\eta)\, ,
\eea
where 
\bea\label{eq:B1Lusual}
  B_{222}(k_1,k_2,k_3,\eta) &=& 8\int_{k_m} d^3q P_0(q)P_0(|\bq+\bk_1|)P_0(|\bq-\bk_2|)F_2(-\bq,\bq+\bk_1,\eta)F_2(-\bq-\bk_1,\bq-\bk_2,\eta)\nn\\
  && \times F_2(\bk_2-\bq,\bq,\eta)\, ,\nn\\
  B_{321}^I(k_1,k_2,k_3,\eta) &=& 6P_0(k_3)F_1(k_3,\eta)\int_{k_m} d^3q P_0(q)P_0(|\bq-\bk_2|)F_3(-\bq,\bq-\bk_2,-\bk_3,\eta)F_2(\bk_2-\bq,\bq,\eta)\nn\\
  && + 5 \ {\rm permutations}\, ,\nn\\
  B_{321}^{II}(k_1,k_2,k_3,\eta) &=& 6P_0(k_2)P_0(k_3)F_2(\bk_2,\bk_3,\eta)F_1(k_2,\eta)\int_{k_m} d^3q P_0(q)F_3(\bk_3,\bq,-\bq,\eta)\nn\\
  && + 5 \ {\rm permutations}\, ,\nn\\
  B_{411}(k_1,k_2,k_3,\eta) &=& 12P_0(k_2)P_0(k_3)F_1(k_2,\eta)F_1(k_3,\eta)\int_{k_m} d^3q P_0(q)F_4(\bq,-\bq,-\bk_2,-\bk_3,\eta)\nn\\
  && + 2 \ {\rm permutations}\, .
\eea
Compared to SPT, the EFT kernels $F_n\equiv F_{n,1}$ depend nontrivially on time $\eta$, as indicated, as well as on $k_m$.
As discussed above, even the linear solution described by $n=1$ has a nontrivial time and scale dependence. For this reason we
explicitly include also the kernel $F_1(k,\eta)$ in the expressions above. In addition, the cutoff at $k_m$ is indicated by the
subscript on the loop integrals. Finally, $P_0(k)$ denotes the usual linear power spectrum as obtained from CLASS \cite{Blas:2011rf}.

Within the viscous EFT approach, we compute the bispectrum numerically
 as a sum of three terms:
\be\label{eq:bi_eft}
  B^{EFT}(k_1,k_2,k_3) = B_{\rm tree}^{EFT}(k_1,k_2,k_3)+B_{\rm one-loop}^{EFT}(k_1,k_2,k_3)+B_{\rm vertex}^{EFT}(k_1,k_2,k_3).
\ee
The first and second term correspond to the tree-level and one-loop contributions computed as described above, with
time-dependent kernels $F_n$ evaluated numerically through the differential equation \eqref{kernels}. 
For the one-loop integration we combine the contributions within the loop integrand in order to achieve the
cancellation of infrared singularities at the integrand level, 
as described for the power spectrum in \cite{Blas1}. We explain the details
of this process in appendix \ref{appendixb}.
In the numerical solution we take into account 
the effective viscosity and pressure contained in $\Omega_{ab}(\bk,\eta)$ exactly, while we use only the SPT contributions to the vertices $\Gamma_{abc}\to\gamma_{abc}$ on the right-hand side.
In order to take the leading contribution of the vertex correction into account, we compute in addition the tree-level bispectrum with a
vertex $\delta\gamma_{abc}=\Gamma_{abc}-\gamma_{abc}$ and viscous propagators, denoted by $B_{\rm vertex}^{EFT}$. Our numerical results
indicate that the vertex correction gives only a minor contribution (see below), justifying the expansion in $\delta\gamma_{abc}$. This conclusion is 
consistent with the discussion at the end of section \ref{recurrenc}.

The numerical results shown below correspond to a $\Lambda$CDM reference cosmology with parameters $\Omega_\Lambda=0.74, \Omega_m=0.26, \Omega_b=0.044, h=0.72, n_s=0.96$.
The quantity $\sigma_d^2$, determining the effective viscosity, pressure and vertices, is computed
at $z=0$ and takes the values 
$\sigma_d^2=\{2.35, 1.37, 0.92, 0.66\}({\rm Mpc}/h)^2$ 
for $k_m=\{0.4, 0.6, 0.8, 1.0\}\, h/$Mpc. For the main analysis we use
 $\lambda_\nu=38/35\,\sigma_d^2$ and $\lambda_s=57/35\,\sigma_d^2$,
as in \cite{viscousdm}, but we also examine the dependence on 
the ratio $\lambda_\nu/\lambda_s$.

\subsection{Dependence on the renormalization scale and comparison with SPT}

As a cross-check of the validity of the viscous EFT description, it is instructive 
to consider the consistency with SPT in the appropriate approximation, as well as the dependence on the renormalization scale $k_m$.
Within the effective theory, $B_{\rm tree}^{EFT}$ depends on the effective viscosity and pressure, parameterized by $\lambda_\nu(k_m)$ and $\lambda_s(k_m)$, which 
are running with $k_m$.
Therefore, $B_{\rm tree}^{EFT}$ depends on $k_m$ as well.
In addition, the contribution $B_{\rm vertex}^{EFT}$ depends on $k_m$ via the running of the effective vertex $\delta\gamma_{abc}(k_m)$.

The running of the EFT parameters is determined by ``integrating out'' UV modes. In general, the running deep inside the nonlinear regime cannot be predicted perturbatively,
but has to be extracted from numerical simulation data. However, the dominant contribution to the effective viscosity and pressure is arguably generated when integrating out modes
$k\gtrsim k_m$ that are not too far away from the weakly nonlinear regime. Independently of the accuracy with which this expectation is borne out, the matching prescription
followed in this work implies that the sum
\be
  B_0 \equiv B_{\rm tree}^{EFT}(k_1,k_2,k_3;k_m)+B_{\rm vertex}^{EFT}(k_1,k_2,k_3;k_m)+B_{\rm one-loop}^{SPT,\ \Lambda=k_m}(k_1,k_2,k_3)
\ee
should be approximately independent of $k_m$, and equal to the usual SPT bispectrum computed using EdS kernels $F_n$:
\be\label{eq:Bspt}
  B^{SPT}(k_1,k_2,k_3) = B_{\rm tree}^{SPT}(k_1,k_2,k_3)+B_{\rm one-loop}^{SPT}(k_1,k_2,k_3)\,.
\ee
The terms $B_{\rm tree}^{EFT}(k_1,k_2,k_3;k_m)$ and 
$B_{\rm vertex}^{EFT}(k_1,k_2,k_3;k_m)$
were defined in the paragraph below eq. (\ref{eq:bi_eft}).
The term $B_{\rm one-loop}^{SPT,\ \Lambda}$ denotes the
SPT one-loop contribution to the bispectrum, computed with the usual SPT kernels and a UV cutoff $\Lambda$.
A small difference $B_0-B^{SPT}$ implies that the tree-level 
parameters contained in the 
EFT are sufficient to capture the UV one-loop contributions 
in the context of SPT within the weakly nonlinear regime.
The confirmation that this expectation is indeed fulfilled constitutes a nontrivial check of the validity range of the EFT description.

\begin{figure}[t]
\centering
\includegraphics[width=0.95\textwidth]{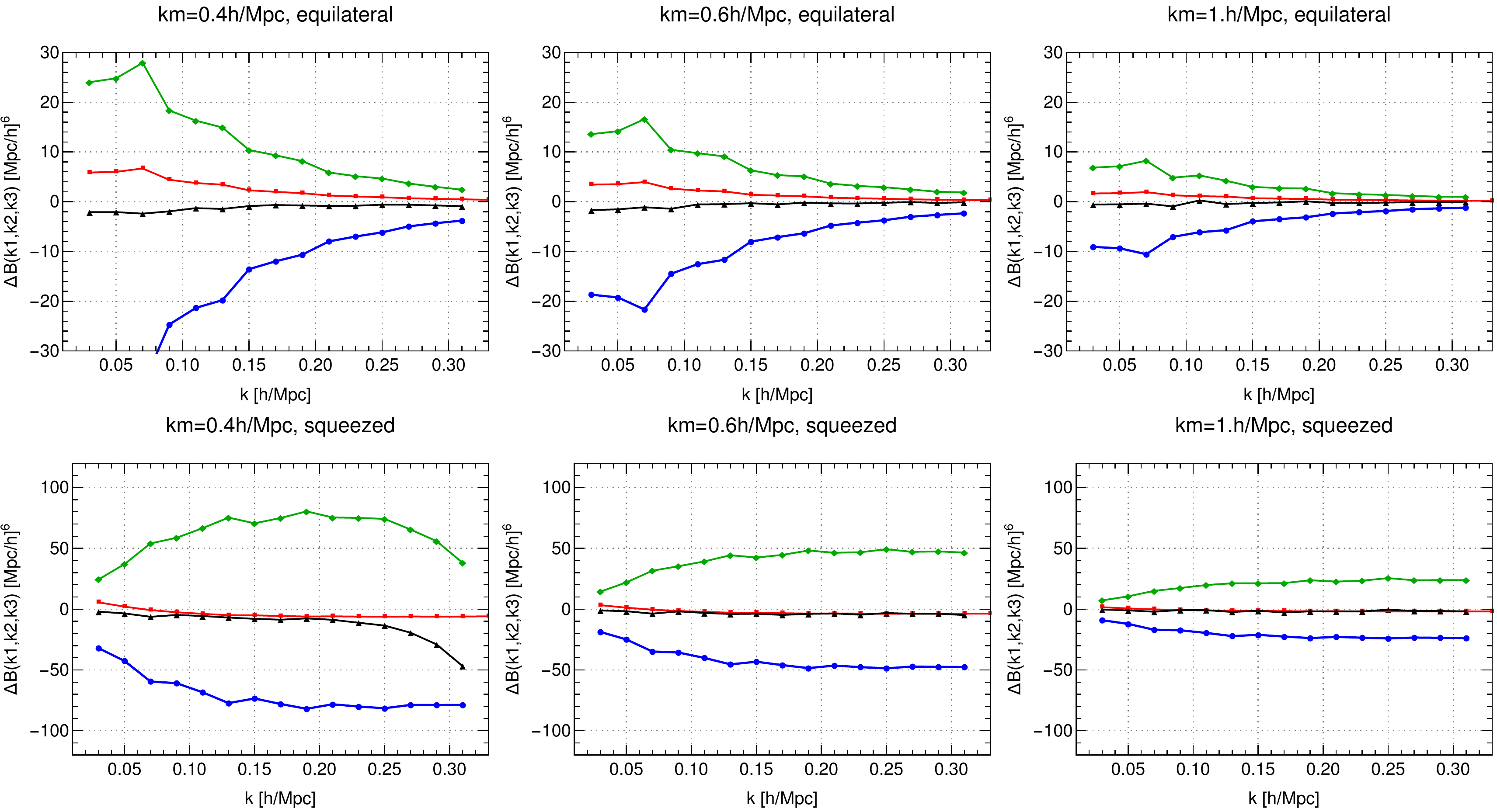}
\caption{
Cross check of the validity of the EFT description, for equilateral configurations $k_1=k_2=k_3\equiv k$ (upper row) 
and for a squeezed shape with $k_1=k_2\equiv k$ and $k_3=0.027\, h/$Mpc (lower row).
The three columns show results for three values of the renormalization scale $k_m$. 
In each panel, the blue and red lines correspond to $B_{\rm tree}^{EFT}-B_{\rm tree}^{SPT}$ and  $B_{\rm vertex}^{EFT}$, respectively. The green line corresponds to
$B_{\rm one-loop}^{SPT, \ \Lambda=k_m}-B_{\rm one-loop}^{SPT}$, where the former denotes the SPT one-loop bispectrum computed with a UV cutoff $\Lambda\equiv k_m$.
The black line corresponds to the sum of the green, blue and red lines, and is expected to be close to zero within the range of validity of the EFT description,
i.e. whenever the effective viscosity and pressure as
well as the effective vertices accurately capture the UV dependence of the bispectrum.
}
\label{fig:bi1}
\end{figure}

In Fig.\,\ref{fig:bi1} we show $B_0-B^{SPT}$ for three values of $k_m$, and for equilateral as well as squeezed configurations, respectively (black lines).
We find that the difference is indeed very close to zero in all cases. One exception is the regime $k\gtrsim 0.25\, h/$Mpc for the squeezed shape and for
the smallest renormalization scale $k_m=0.4\, h/$Mpc. One reason for this might be that the EFT is formally valid for the limit of a large scale separation, $k\ll k_m$.
Therefore, the regime in which $k$ and $k_m$ approach each other is precisely where the EFT description is expected to start breaking down.

It is also instructive to decompose $B_0-B^{SPT}$ as 
\be
  B_0-B^{SPT} = \left(  B_{\rm tree}^{EFT}-B_{\rm tree}^{SPT}\right) + B_{\rm vertex}^{EFT} + \left(B_{\rm one-loop}^{SPT,\ \Lambda=k_m}-B_{\rm one-loop}^{SPT}\right).
\ee 
The three contributions on the right-hand side are shown in Fig.\,\ref{fig:bi1} by the blue, red and green lines, respectively.
As expected, the individual contributions depend on $k_m$, and the dependence on the renormalization scale only cancels when adding all contributions.
This indicates that the running EFT parameters indeed capture the UV contributions
efficiently. 

The smallness of the contribution $B_{\rm vertex}^{EFT}$ 
relative to the other two supports the conclusion we reached at the end of
section \ref{recurrenc}, that 
the calculation of higher-order kernels 
within the effective theory can be performed by keeping the full 
matrix $\Omega_{ab}$ of eq. (\ref{omeren}), but
neglecting the vertex corrections $\delta\gamma_{abc}$.
On the other hand, $B_{\rm vertex}^{EFT}$ is not negligible at tree-level.
From the first row in Fig.\,\ref{fig:bi1} it is apparent that the vertex correction 
(red line) gives a small but still significant contribution
for the equilateral configuration. On the other hand, the vertex correction is not required in order to capture the UV sensitivity of the bispectrum in the
squeezed limit, to a good approximation. This implies that the bispectrum in the squeezed limit can be described by a simplified EFT that contains only an
effective viscosity and pressure, but no vertex corrections. Moreover, effectively only the linear combination $\lambda_\nu(k_m)+\lambda_s(k_m)$ enters in
the viscous propagator when $k\lesssim k_m$. This suggests that the viscous description employed in \cite{viscousdm}, with a single EFT parameter, is sufficient to predict both the
power spectrum as well as the bispectrum in the squeezed limit, while an additional parameter (for the vertex correction) is required to achieve percent precision 
in the equilateral case. We leave further investigation of this point for future work. This finding is also broadly in agreement with results based on different types of effective theory constructions \cite{Baldauf:2014qfa,Angulo:2014tfa}.

\begin{figure}[t]
\centering
\includegraphics[width=0.95\textwidth]{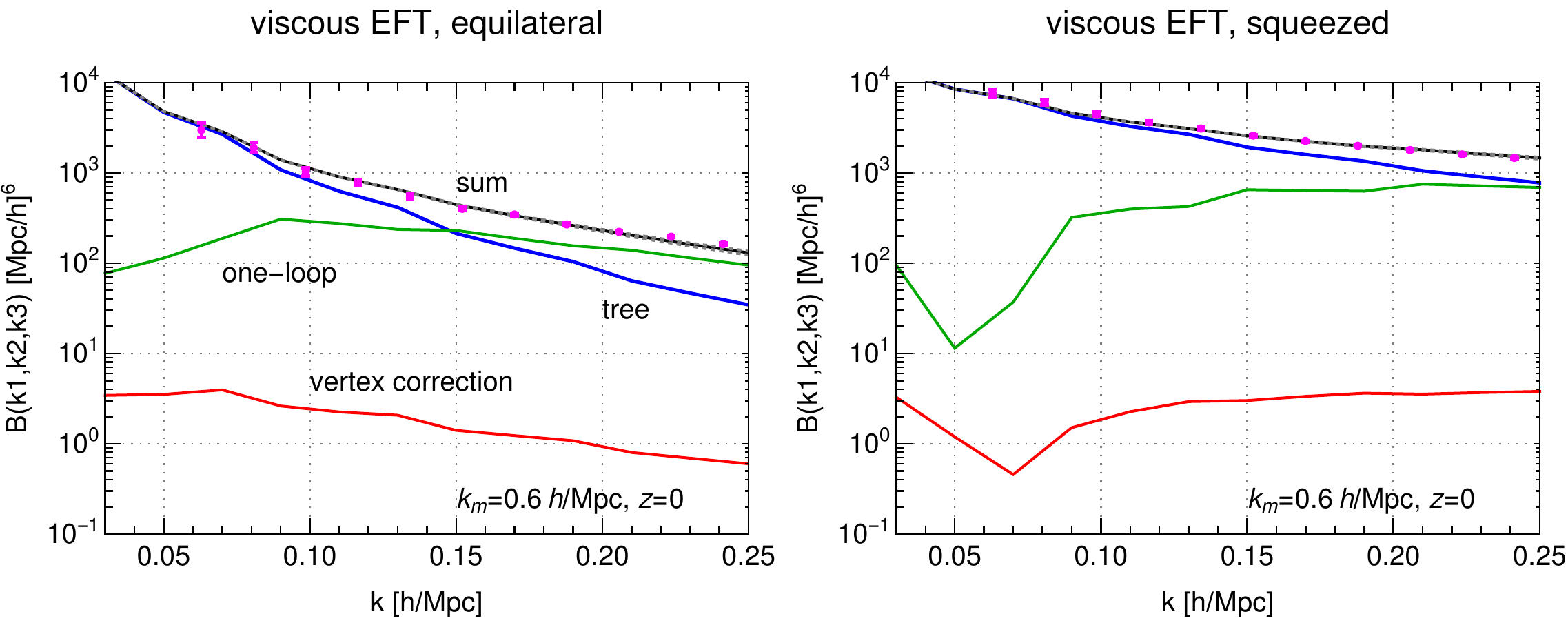}
\caption{
Bispectrum for equilateral configurations $k_1=k_2=k_3\equiv k$ (left) and for a squeezed shape with $k_1=k_2\equiv k$ and $k_3=0.027\, h/$Mpc (right).
The black line shows the result for the total
bispectrum, and the blue and green correspond to the tree-level and one-loop contributions, respectively. For the viscous EFT the additional contribution due
to vertex corrections is shown in red. The grey dotted lines show the variation with the renormalization scale in the range $k_m=0.4-1\, h/$Mpc (they are
almost indistinguishable from the black line). The data-points (in magenta) correspond to N-body simulation results \cite{Baldauf:2014qfa}.
}
\label{fig:bilog}
\end{figure}

\begin{figure}[t]
\centering
\includegraphics[width=0.95\textwidth]{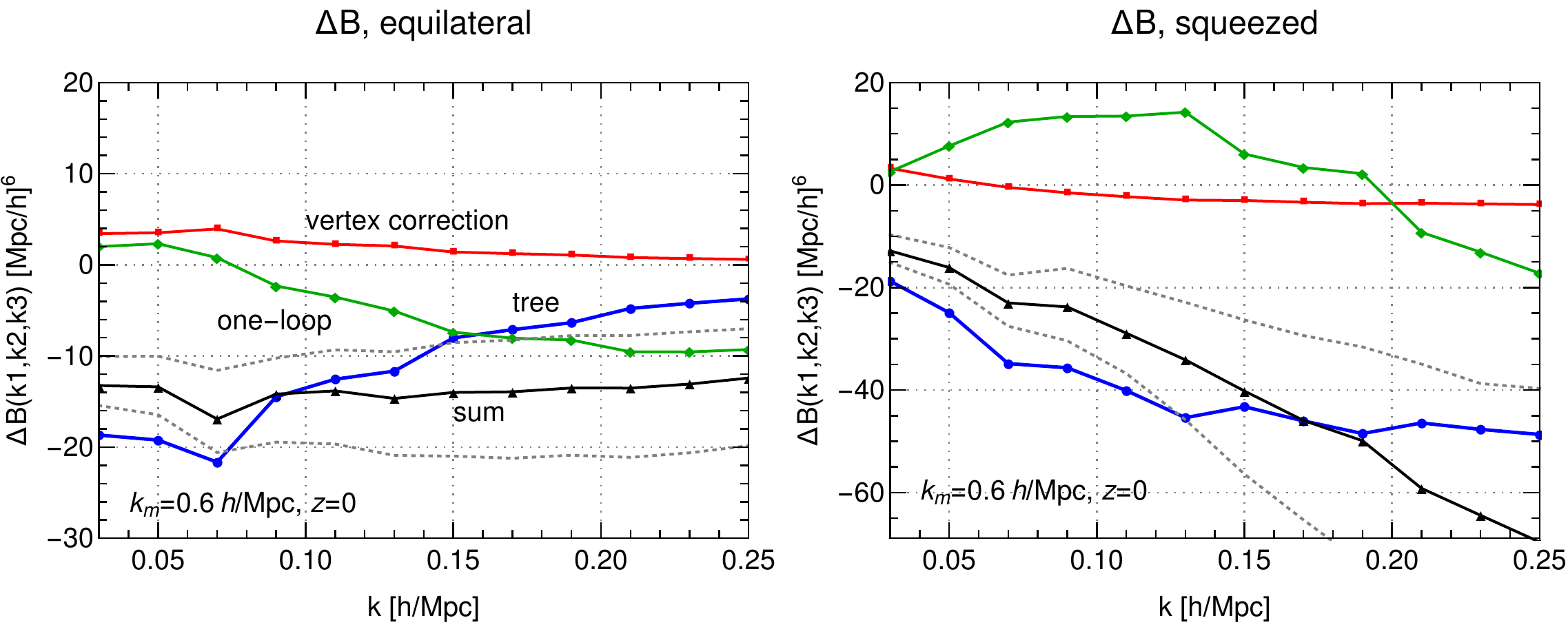}
\caption{
Difference between the bispectrum computed in the viscous EFT approach and in SPT, for equilateral and squeezed configurations as in Fig.\,\ref{fig:bilog},
and for $k_m=0.6\, h/$Mpc (black lines). For the tree-level contribution (blue) we show $B_{\rm tree}^{EFT}-B_{\rm tree}^{SPT}$, while for
the loop correction (green) we display $B_{\rm one-loop}^{EFT}-B_{\rm one-loop}^{SPT}$. The vertex correction (red) corresponds to $B_{\rm vertex}^{EFT}$, since this contribution does not exist within SPT.
The grey dotted lines show the variation with the renormalization scale in the range $k_m=0.4-1\, h/$Mpc, providing a quantitative measure for
the theoretical uncertainty band.
}
\label{fig:dbi}
\end{figure}

\begin{figure}[t]
\centering
\includegraphics[width=0.45\textwidth]{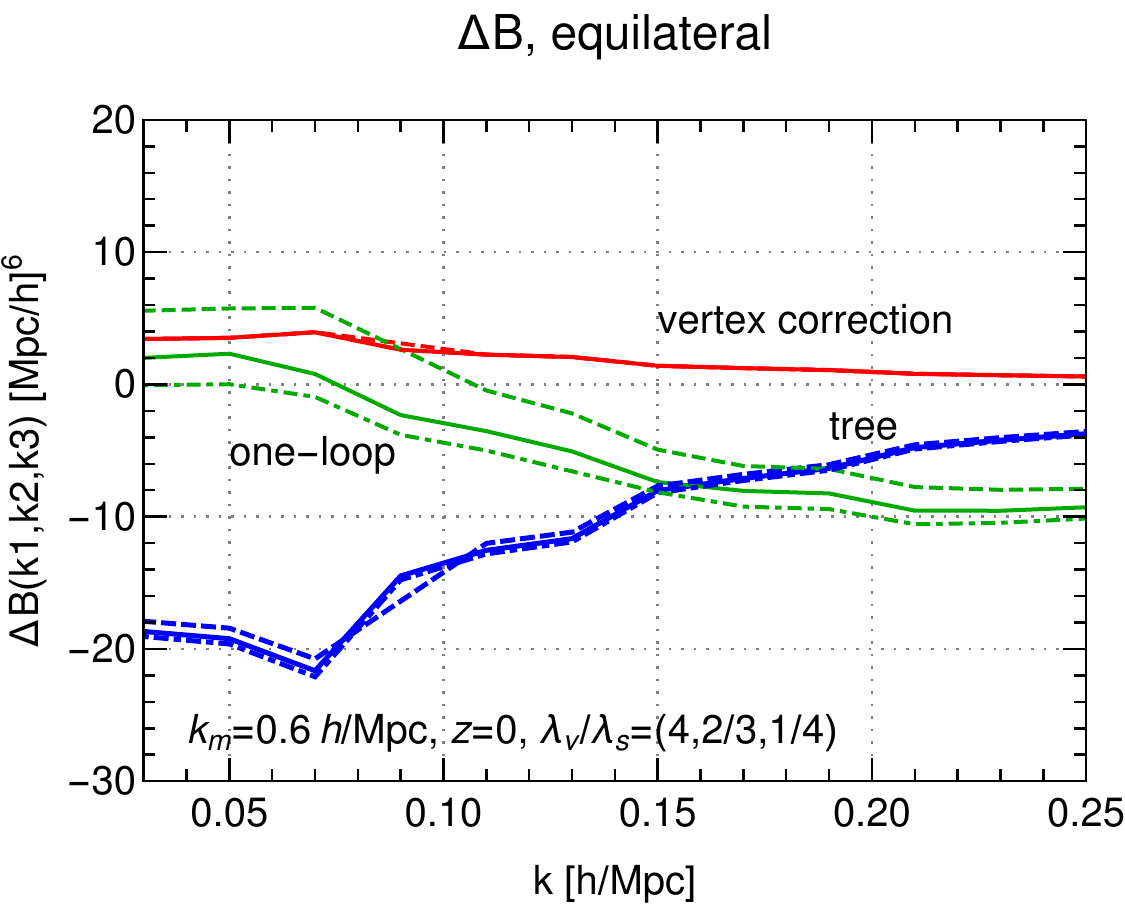}
\includegraphics[width=0.45\textwidth]{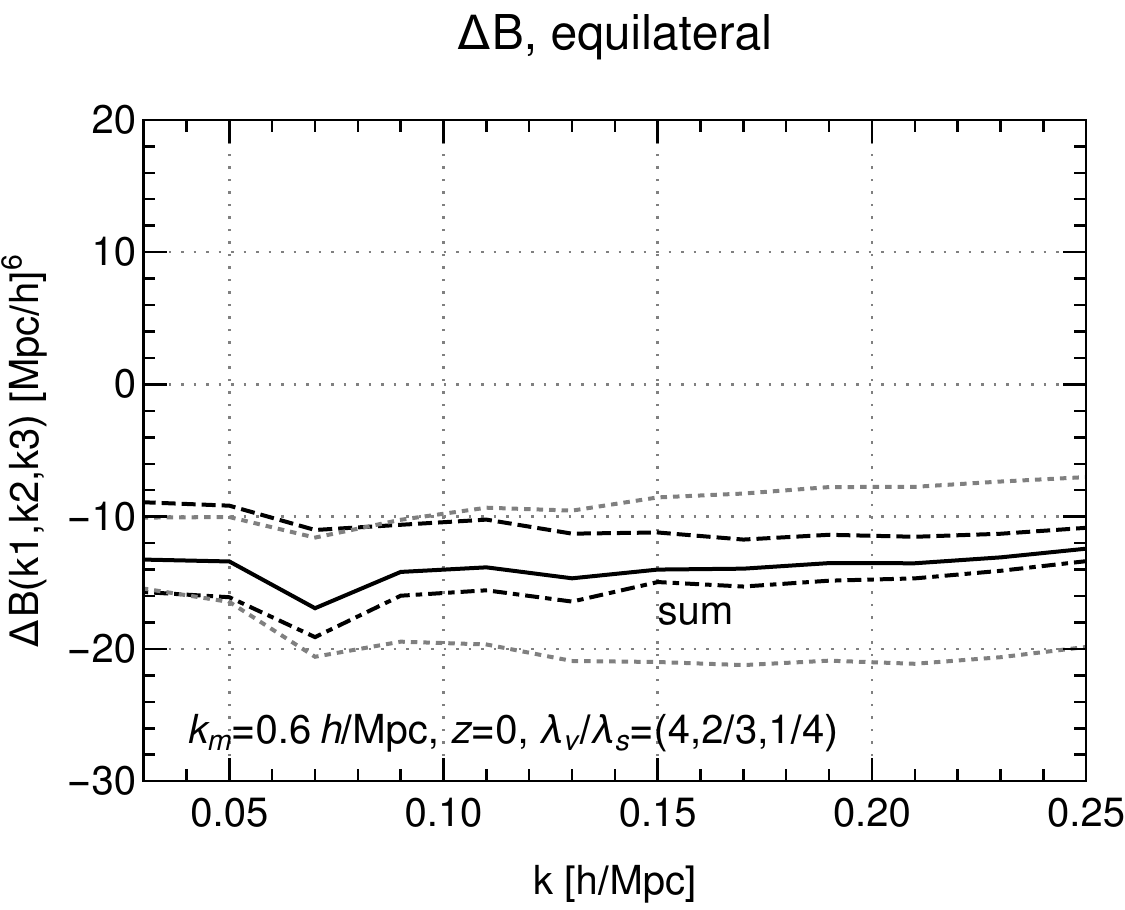}
\caption{
Dependence of the bispectrum on the relative contribution of effective viscosity $\bar\lambda_\nu$ and sound velocity $\bar\lambda_s$, for $k_m=0.6\, h/$Mpc and equilateral shape.
Solid lines correspond to the fiducial choice $\bar\lambda_\nu/\bar\lambda_s=2/3$ \cite{viscousdm}. The result obtained for an identical value of 
the sum $\bar\lambda_\nu+\bar\lambda_s$, but with $\bar\lambda_\nu/\bar\lambda_s=4$, is shown by the dashed lines, while dot-dashed lines correspond to $\bar\lambda_\nu/\bar\lambda_s=1/4$.
The left panel shows the individual contributions from tree-level (blue), one-loop (green) and vertex correction (red) as in Fig.\,\ref{fig:dbi}, and the right
panel contains their sum. For comparison, the grey dotted lines in the right panel show the dependence of the bispectrum on $k_m$ for the fiducial value
of $\bar\lambda_\nu/\bar\lambda_s$, as in Fig.\,\ref{fig:dbi}.
}
\label{fig:alphasVsnu}
\end{figure}

\subsection{Bispectrum within the effective theory}

We turn next to the calculation of the one-loop bispectrum within the effective theory.
As seen in Fig.\,\ref{fig:bilog}, the viscous EFT description compares well with N-body simulation 
results~\cite{Baldauf:2014qfa}, and vertex corrections indeed make only a
minor contribution in both, equilateral and squeezed configurations.

Fig.\,\ref{fig:dbi} displays the difference between the EFT and the corresponding SPT results.
For the SPT result we do not impose any cutoff in the loop integration, and we use the usual EdS Kernels $F_n$ (see eq.~(\ref{eq:Bspt})).
The differences range from a few percent to several tens of percent for large $k$.
The difference in the tree-level (blue) and one-loop (green) results can be associated with the effective viscosity and sound velocity within the EFT.
As observed before, its effect is subleading to the effective viscosity and sound velocity.

In Figs.\,\ref{fig:bilog} and \,\ref{fig:dbi}, we have included theoretical uncertainty bands, obtained from varying the renormalization scale $k_m$. 
In Fig.\,\ref{fig:bilog}, these are hardly distinguishable from the black line, and they amount to an uncertainty on the percent level below the
nonlinear scale (around $0.2\, h/$Mpc for $z=0$). In Fig.\,\ref{fig:dbi} they are better visible due to the much smaller range of the $y$-axis.
An additional uncertainty, not included in these error bands, arises from two- and higher loop contributions, which we expect to contribute significantly
for $k\gtrsim 0.2\, h/$Mpc.

\begin{figure}[t]
\centering
\includegraphics[width=0.95\textwidth]{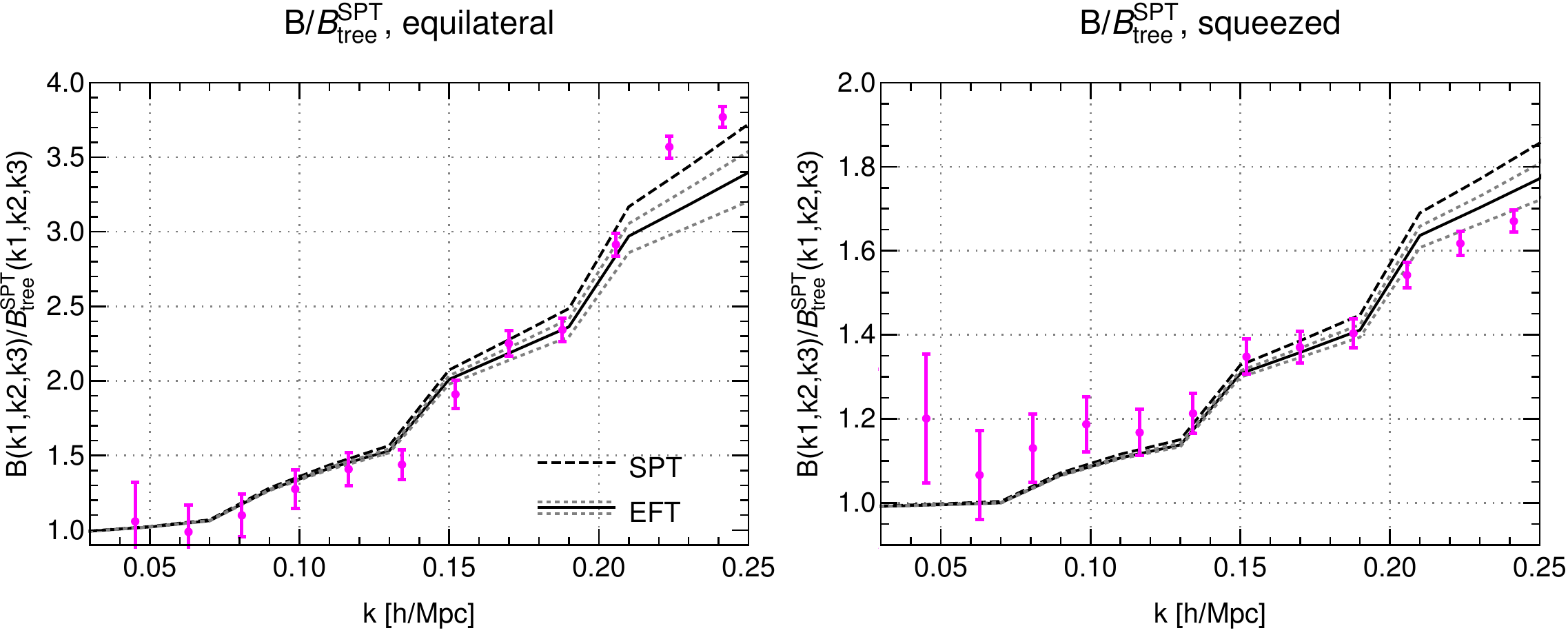}
\caption{
Ratio of the SPT (black dashed) and EFT (black solid) bispectrum to the SPT tree-level bispectrum.
The grey dotted lines show the variation with the renormalization scale in the range $k_m=0.4-1\, h/$Mpc, providing a quantitative measure for
the theoretical uncertainty band below the nonlinear scale ($k\lesssim 0.2\, h/$Mpc). 
The data-points (in magenta) correspond to N-body simulation results \cite{Baldauf:2014qfa}.
}
\label{fig:rbi}
\end{figure}

An additional source of systematic uncertainty arises from the relative size of effective viscosity  
$\bar\lambda_\nu$ and sound velocity $\bar\lambda_s$. For the matching procedure discussed in section \ref{sec2}, only their sum is determined. In Fig.\,\ref{fig:alphasVsnu}
we quantify the dependence of the bispectrum on the ratio $\bar\lambda_\nu/\bar\lambda_s$, within the range $1/4-4$, including also the fiducial
value $\bar\lambda_\nu/\bar\lambda_s=2/3$ \cite{viscousdm}. The left panel of Fig.\,\ref{fig:alphasVsnu} demonstrates that the one-loop contribution to the
bispectrum shows a mild dependence on $\bar\lambda_\nu/\bar\lambda_s$, while both the tree-level and vertex contributions are almost insensitive to this
parameter. The variation of the total bispectrum (black lines in the right panel of Fig.\,\ref{fig:alphasVsnu}) when changing 
$\bar\lambda_\nu/\bar\lambda_s$ within the range $1/4-4$ is comparable to (or smaller than) the dependence on the value of $k_m$ (grey dotted lines).
Therefore, the latter provides the dominant contribution to the theoretical error budget. 

Fig.\,\ref{fig:rbi} shows the ratio of the SPT and EFT bispectra to the tree-level SPT bispectrum, compared to 
N-body simulation results \cite{Baldauf:2014qfa}. 
We find good agreement within the expected range of validity ($k\lesssim 0.2\, h/$Mpc). For $k\lesssim 0.15 \, h/$Mpc, the N-body simulation results are
limited by cosmic variance, and do not lead to a meaningful distinction 
between SPT and EFT. For  $k\gtrsim 0.2\, h/$Mpc, the validity of the 
perturbative description is expected to become weaker at $z=0$
in comparison to N-body data. 
Within the range $0.15 \, h/$Mpc $\lesssim k \lesssim 0.2\, h/$Mpc,
the EFT results are in slightly better agreement with $N$-body data 
than the SPT results, even though the difference is not substantial. 
We emphasize that
no parameters have been adjusted in order to obtain the 
EFT results for the bispectrum.
We thus conclude that the viscous EFT provides---without inclusion of additional effective vertex corrections---a stable and reliable description of both spectra and bispectra for sufficiently large wavelength ($k<0.2\, h/$Mpc).

\section{Conclusions} \label{conclusions}

In this work we continued the program of constructing a one-particle 
irreducible effective action $\Gamma_{k_m}[\phi,\chi]$ 
for large-scale structure at a scale $k_m$ below the typical galaxy scale. 
Conceptually, the effective action results from taking into account
the fluctuations at scales 
$|{\bf k}|>k_m$ through the introduction of effective couplings. 
This effective action can be 
taken as a starting point for computing additional loop corrections,
now in the presence of a regulator restricting the integrations to 
$|{\bf k}|< k_m$, or for calculating the 
functional RG running towards the full one-particle irreducible effective action $\Gamma[\phi,\chi]$ \cite{Floerchinger:2016hja}. We emphasize that in a one-particle irreducible scheme the effective propagator 
derived from $\Gamma_{k_m}[\phi,\chi]$ receives corrections at all orders in a perturbative expansion 
(by virtue of the one-particle irreducible resummation). 
Such a resummed propagator can be obtained by 
solving equations of motion with ``self-energy'' corrections, 
as was discussed in section \ref{sec2}. We take this resummation into account
when computing loop corrections within the effective theory for $|{\bf k}|< k_m$.

The RG running to the scale $k_m$ generates
effective viscosity and sound velocity terms in $\Gamma_{k_m}[\phi,\chi]$, even when starting with an ideal pressureless 
fluid description in the ultraviolet. The resulting renormalization group equations indicate that, for a realistic spectrum
of perturbations, the dominant contributions to the effective viscosity and sound velocity arise from scales that are only slightly above
$k_m$ \cite{viscousdm, Floerchinger:2016hja}. On the one hand, this finding is in line with the expected decoupling of UV modes.
On the other hand, the starting
point of the RG evolution in the UV may deviate from a pressureless ideal fluid
on general gounds. In such a case the RG evolution of
the effective viscosity and sound velocity would start from (in general unknown) non-zero values in the UV.
We have found previously that the power spectra computed in the 
absence of such extra UV contributions are approximately independent 
of the artificial intermediate scale $k_m$ and
in reasonable agreement with N-body simulations. This may be taken as an indication that the (computable) RG running within the
perturbative domain captures the dominant contribution to the effective viscosity and sound velocity evaluated at the scale $k_m$.

Here we extended this analysis to the bispectrum. In addition to effective viscosity and sound velocity
terms, we also included vertex corrections in the effective action $\Gamma_{k_m}[\phi,\chi]$ that are 
generated by integrating out fluctuations at scales $|{\bf k}|>k_m$. In order to estimate the importance of
vertex corrections, we limited ouselves to a perturbative determination instead of a self-consistent RG analysis. 
We showed that the resulting
effective theory accurately captures the impact of UV fluctuations on the bispectrum, provided that
the scale $k_m$ lies within the regime where a perturbative description is possible.
As for the power spectrum, the resulting bispectrum depends only mildly on $k_m$.
While the vertex corrections are relevant in order to achieve a precise cancellation of the dependence
on $k_m$, their quantitative contribution to the total bispectrum is of minor importance, especially
for squeezed configurations. This implies that the dominant effect is captured by the effective propagator,
and is consistent with the truncation of the effective action used in \cite{viscousdm, Floerchinger:2016hja}.

The results lend further support to the viscous fluid description based on the one-particle 
irreducible effective action $\Gamma_{k_m}[\phi,\chi]$. In future work the framework could be further developed, for example by including additional fields such as vorticity or the velocity dispersion tensor \cite{McDonald:2009hs, Erschfeld:2018zqg}, as well as by using the functional renormalization group flow for computations at small wavenumbers $k < k_m$.

\section*{Acknowledgments}

S.\ F.\ and M.\ G.\ acknowledge support by the Munich Institute for Astro- and Particle Physics (MIAPP) of the DFG cluster of excellence ORIGINS.
The work of S.\ F.\ is supported by Deutsche Forschungsgemeinschaft (DFG) under EXC-2181/1-390900948 (the Heidelberg STRUCTURES Excellence Cluster) as well as BE 2795/4-1.

\appendix

\section{Calculation of the vertex correction} \label{appendixa}

The vertex correction is obtained by computing one-particle-irreducible (1PI) diagrams with two ingoing
lines and one outgoing line, such as the ones shown in Fig.\,\ref{fig2}.

At one-loop, we take three diagrams into account. The first one is shown on the right-hand side
in Fig.\,\ref{fig2} (which we denote by diagram ($1$)). The second one (diagram ($2$)) has a similar structure,
except that the initial power spectrum (depicted by the small square) is inserted in the internal
line connecting $\eta_1$ and $\eta_3$. In the third diagram ($3$), the square is inserted in the line connecting $\eta_2$ and $\eta_3$.

We first compute the amputated diagrams, i.e. without propagators attached to the three external lines. Denoting the incoming wavenumber
at $\eta_1$ by $\bp$ and at $\eta_2$ by $\bq$, they are given by
\bea
  V_{abc}^{amp}(\eta_1,\eta_2,\eta_3,\bp,\bq)&=& 8\int d^3l\,P_0(l)\,\Big(V_{abc}^{(1)}\Theta(\eta_3-\eta_1)\Theta(\eta_3-\eta_2)+V_{abc}^{(2)}\Theta(\eta_3-\eta_2)\Theta(\eta_2-\eta_1)\nn\\
  && +V_{abc}^{(3)}\Theta(\eta_3-\eta_1)\Theta(\eta_1-\eta_2)\Big)\, ,
\eea
where we explicitly extracted the initial, linear power spectrum $P_0(l)$ that corresponds to the ``square'' in the internal lines,
and the Heaviside functions associated to the propagator. The factor $8=2^3$ is related to the combinatorial factor for each of the three vertices.
The loop integrands for the three diagrams are given by
\bea
  V_{abc}^{(1)} &=& \gamma_{add'}(\bp+\bl,\bq-\bl)\left(g^R_{de}(\eta_3-\eta_1)\gamma_{ebf}(\bp,\bl)\right)\left(g^R_{d'e'}(\eta_3-\eta_2)\gamma_{e'cf'}(\bq,\bl)\right)\left(u_fu_{f'}e^{\eta_1+\eta_2}\right)\, ,\nn\\
  V_{abc}^{(2)} &=& \gamma_{add'}(-\bl,\bp+\bq+\bl)\left(u_du_{e}e^{\eta_3+\eta_1}\right)\left(g^R_{d'e'}(\eta_3-\eta_2)\gamma_{e'cf'}(\bq,\bp+\bl)\right)\left(g^R_{f'f}(\eta_2-\eta_1)\gamma_{fbe}(\bp,\bl)\right)\, , \nn\\
  V_{abc}^{(3)} &=& V^{(2)}\big|_{\eta_1\leftrightarrow\eta_2,\bp\leftrightarrow\bq}\, ,
\eea
where summation over repeated indices is implied and $u=(u_1,u_2)=(1,1)$ projects out the growing mode contribution
for the propagators attached to the initial power spectrum.

We are interested in the UV contribution from modes with $l\geq k_m\gg p,q$. We therefore Taylor expand the $V^{(i)}$ in $1/l$.
We find cancellations among the three contributions, and the leading large-$l$ behaviour can be extracted by rewriting
the vertex in the form
\bea\label{eq:Vamp}
  V_{abc}^{amp,UV}(\eta_1,\eta_2,\eta_3,\bp,\bq)&=& 8\Theta(\eta_3-\eta_1)\Theta(\eta_3-\eta_2)\int_{k_m}^\infty dl\,l^2\,P_0(l)\,\Big((\bar V_{abc}^{(1)}+\bar V_{abc}^{(2)})\Theta(\eta_2-\eta_1)\nn\\
  && +(\bar V_{abc}^{(1)}+\bar V_{abc}^{(3)})\Theta(\eta_1-\eta_2)\Big)\, ,
\eea
where
\be
  \bar V_{abc}^{(i)} \equiv \int d\Omega_l V_{abc}^{(i)}
\ee
are integrated over the direction of $\bl$. We find that both $\bar V_{abc}^{(1)}+\bar V_{abc}^{(2)}$ and $\bar V_{abc}^{(1)}+\bar V_{abc}^{(3)}$ scale $\propto 1/l^2$ when Taylor expanded
for large $l$. Therefore, the integration on the right-hand side of \eqref{eq:Vamp} yields a factor $\sigma_d^2=4\pi/3 \int_{k_m}^\infty dl P_0(l)$.

The amputated vertex depends separately an all time arguments. In order to match the UV part of the one-loop correction to a modified vertex within the low-energy effective theory,
we therefore consider the non-amputated vertex given by
\be
  V_{abc}^{UV}(\eta,\bp,\bq)= \int_{-\infty}^\eta d\eta_3 \int_{-\infty}^{\eta_3}d\eta_1\int_{-\infty}^{\eta_3}d\eta_2\, g^R_{aa'}(\eta-\eta_3)V_{a'b'c'}^{amp}(\eta_1,\eta_2,\eta_3,\bp,\bq)g^R_{b'b}(\eta_1)g^R_{c'c}(\eta_2)\, ,
\ee
where we project on the growing mode contribution for the incoming lines (attached to $\eta_1$ and $\eta_2$) by sending the initial time to $-\infty$.
For the leading contribution in the Taylor expansion for $p,q\ll k_m$ we find
\bea
  V_{1bc}^{UV}(\eta,\bp,\bq) &=& - \frac{e^{4\eta}\sigma_d^2 v_bv_c }{679140p^2q^2}\Big(22625 k^6 + 37666 k^4 (p^2 + q^2) + 62973 (p^2 - q^2)^2 (p^2 + q^2) \nn\\
  && -  12 k^2 (10272 p^4 - 5255 p^2 q^2 + 10272 q^4)\Big)\, ,\nn\\
  V_{2bc}^{UV}(\eta,\bp,\bq) &=& - \frac{e^{4\eta}\sigma_d^2 v_bv_c }{679140p^2q^2}\Big(26535 k^6 + 91352 k^4 (p^2 + q^2) + 20991 (p^2 - q^2)^2 (p^2 + q^2)  \nn\\
  && -  2 k^2 (69439 p^4 - 263840 p^2 q^2 + 69439 q^4)\Big)\, ,
\eea
where $\bk=\bp+\bq$ and $v=(v_1,v_2)=(3/5,2/5)$.

As a cross check we verified that when adding to this result the 1PR contributions (three diagrams with ``self-energy'' insertions on either of the external lines),
we recover the corresponding SPT result
\be
  V_{1bc}^{SPT}=12v_bv_c\int d^3l \, P_0(l) F_4(\bp,\bq,\bl,-\bl),
\ee
where $F_4$ is the standard, fully symmetrized SPT kernel. Since, within the effective theory, the 1PR contributions are already (approximately) taken into account via the
viscosity and sound velocity corrections to the propagator, we only use the 1PI contributions (as computed above) for the matching to the effective vertices.

Within the effective theory, the corresponding tree-level diagram that involves the
correction $\delta\gamma_{abc}$ to the SPT vertices is given by
\be
  \delta V_{abc}^{EFT}(\eta,\bp,\bq)= 2\int_{-\infty}^\eta d\eta'g^R_{aa'}(\eta-\eta')\delta\gamma_{a'b'c'}(\bp,\bq,\eta')g^R_{b'b}(\eta')g^R_{c'c}(\eta')\, .
\ee
As matching condition, we require agreement of both expressions,
\be
  V_{abc}^{UV}(\eta,\bp,\bq)= \delta V_{abc}^{EFT}(\eta,\bp,\bq).
\ee
This yields two independent relations for two linear combinations of $\delta\gamma_{abc}$, given in \eqref{dg1} and \eqref{dg2}.

It should be noted that the matching conditions are not unique. In particular, for $l\gg k_m$ higher-order corrections, as well
as effects that go beyond the fluid description, are relevant. In addition, the nontrivial time dependence implies that the
effective theory can only be expected to capture the dominant effects arising from the growing mode. Within the approach followed here
the theoretical uncertainty due to both of these points can be assessed by the dependence of the result on the choice of the matching scale $k_m$ (see Sec.\,\ref{sec:num}).

\section{Efficient evaluation of the loop integral for the bispectrum}
\label{appendixb}

For the one-loop contribution to the bispectrum, we combine the contributions within the loop integrand in order to achieve the
cancellation of infrared singularities at the integrand level, as described for the power spectrum in \cite{Blas1}.
Even though we will evaluate the bispectrum at most at one-loop order below, we describe an algorithm that eliminates
infrared singularities on the integrand level at any loop order $L\geq 1$ for completeness. This generalizes the one-loop
case discussed in \cite{Baldauf:2014qfa,Angulo:2014tfa} and the two-loop calculation in \cite{Lazanu:2018yae}.

In order to describe the manipulations performed on the loop integrand, it is convenient to introduce a notation
for a general contribution to the bispectrum at any loop order. Each perturbative contribution $B_{ABC}$ contains a product
of three kernels of the form $F_A\times F_B\times F_C$. Each kernel $F_n(\bq_1,\dots,\bq_n)$ can be pictured as a ``blob'' with $n$ ingoing lines, carrying wavenumbers $\bq_i$,
and one outgoing line with momentum given by $\bk_{\rm out}=\sum \bq_i$. We assume that the outgoing line of $F_A$ has wavenumber $\bk_1$,
the one of $F_B$ carries $\bk_2$, and the one of $F_C$ carries $\bk_3=-\bk_1+\bk_2$.

\begin{figure}[t]
\centering
\includegraphics[width=0.5\textwidth]{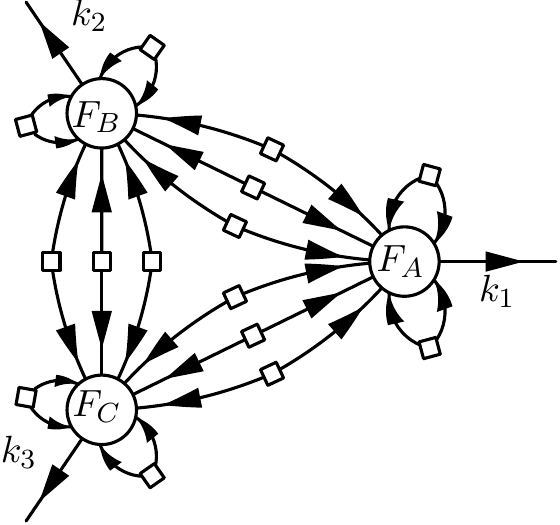}
\caption{Generic structure of an $L$-loop diagram contributing to the bispectrum $B(k_1,k_2,k_3)$. According to the classification discussed in the
text, the diagram has $n_1=n_2=n_3=2$ and $n_{12}=n_{23}=n_{31}=3$. 
}
\label{fig:bi_feyn}
\end{figure}

The most general loop diagram contains internal lines that are attached to a single kernel, and those that connect two kernels.
A generic diagram is schematically shown in Fig.\,\ref{fig:bi_feyn}.
Pairs of two internal lines attached to a single kernel form a loop. In addition, there are loops associated to lines that connect different kernels.
We denote the number of loops formed from lines starting and ending at the same kernel by $n_1$, $n_2$ and $n_3$, for the three
kernels $F_A$, $F_B$, and $F_C$, respectively. The number of lines connecting $F_A$ and $F_B$ is denoted by $n_{12}$, the number of lines connecting $F_B$ and $F_C$ by $n_{23}$,
and the one connecting $F_C$ and $F_A$ by $n_{31}$. The total number of loops is
\be
  L = n_1+n_2+n_3+n_{12}+n_{23}+n_{31}-2\,,
\ee
and the order of the kernels $F_A\times F_B\times F_C$ is given by 
\be
A=2n_1+n_{12}+n_{31},\ B=2n_2+n_{12}+n_{23},\ C=2n_3+n_{23}+n_{31}\,.
\ee
Denoting the corresponding contribution to the bispectrum by $B_{(n_1,n_2,n_3,n_{12},n_{23},n_{31})}$, the familiar one-loop expressions
are given by
\bea
  B_{222} &=& B_{(000111)}\, ,\nn\\
  B_{321}^I &=& B_{(000210)}+B_{(000021)}+B_{(000102)}+B_{(000201)}+B_{(000012)}+B_{(000120)}\, ,\nn\\
  B_{321}^{II} &=& B_{(001110)}+B_{(001101)}+B_{(010101)}+B_{(010011)}+B_{(100110)}+B_{(100011)}\, ,\nn\\
  B_{411} &=& B_{(100101)}+B_{(010110)}+B_{(001011)}\, .
\eea
In each line, the first contribution on the right-hand side corresponds to the one shown explicitly in eq.~\eqref{eq:B1Lusual},
while the other terms contain the permutations.

In this notation, the bispectrum at $L$ loops is
\be
  B_{L-loop}(k_1,k_2,k_3) = \sum_{{n_1,n_2,n_3\geq 0}\atop{n_{12},n_{23},n_{31}\geq 0}}\delta_{L+2,n_1+n_2+n_3+n_{12}+n_{23}+n_{31}}B_{(n_1,n_2,n_3,n_{12},n_{23},n_{31})}(k_1,k_2,k_3)\, ,
\ee
with the additional constraint that at least two out of the indices $(n_{12},n_{23},n_{31})$ are nonzero in order to remove
disconnected pieces.
Each individual contribution is given by
\bea
  B_{(n_1,n_2,n_3,n_{12},n_{23},n_{31})}(k_1,k_2,k_3) &=& \frac{A!B!C!}{2^{n_1+n_2+n_3}n_1!n_2!n_3!n_{12}!n_{23}!n_{31}!}\int dQ\nn\\
  && F_A(q^{(1)}_{1},- q^{(1)}_{1},\dots, q^{(1)}_{n_1},- q^{(1)}_{n_1}, -q^{(12)}_1,\dots, -q^{(12)}_{n_{12}}, q^{(31)}_1,\dots, q^{(31)}_{n_{31}})\nn\\
  && F_B(q^{(2)}_{1},- q^{(2)}_{1},\dots, q^{(2)}_{n_2},- q^{(2)}_{n_2}, -q^{(23)}_1,\dots, -q^{(23)}_{n_{23}}, q^{(12)}_1,\dots, q^{(12)}_{n_{12}})\nn\\
  && F_C(q^{(3)}_{1},- q^{(3)}_{1},\dots, q^{(3)}_{n_3},- q^{(3)}_{n_3}, -q^{(31)}_1,\dots, -q^{(31)}_{n_{31}}, q^{(23)}_1,\dots, q^{(23)}_{n_{23}})\nn\\
\eea
where 
\be\label{eq:dQ}
  \int dQ \equiv \int \prod_{i=1}^{L+2} d^3 Q_i P_0(Q_i)\, \delta^{(3)}\left(\bk_1-\bq_{31}+\bq_{12}\right)
  \delta^{(3)}\left(\bk_2-\bq_{12}+\bq_{23}\right)\, .
\ee
The $\{Q_i\}$ denote collectively the set of $L+2$ wavevectors $\{q^{(1,2,3)}_{j},q^{(12,23,31)}_{k}\}$, and
\be
  \bq_{12}\equiv \sum_{j=1}^{n_{12}} q^{(12)}_j,\
  \bq_{23}\equiv \sum_{j=1}^{n_{23}} q^{(23)}_j,\
  \bq_{31}\equiv \sum_{j=1}^{n_{32}} q^{(31)}_j,\
\ee
denote the total wavevector exchanged between the three ``blobs'', respectively.
Note that the product of the two Dirac functions in \eqref{eq:dQ} implies also that $\bk_3-\bq_{23}+\bq_{31}=0$,
since $\bk_3=-\bk_1-\bk_2$.

We can discriminate two cases: (a) all of the indices $(n_{12},n_{23},n_{31})$ are nonzero, and (b) one of them is zero.

Let us first discuss case (b). At one-loop, this is realized for $B_{321}^{I,II}$ and $B_{411}$. 
We assume as an example that $n_{12}=0$ (the other cases are analogous). This implies that $\bq_{12}=0$, and therefore
the two Dirac functions in \eqref{eq:dQ} fix $\bq_{23}=-\bk_2$ and $\bq_{31}=\bk_1$ in terms of external wavevectors.
This can be used to eliminate the integrations over $Q_{L+2}\equiv q^{(23)}_1$ and $Q_{L+1}\equiv q^{(31)}_1$, so that exactly $L$ loop integrals
over $Q_1,\dots,Q_L$ appear. 
Furthermore, if $n_{23}\geq 2$ (as for $B_{321}^{I}$ at one-loop), the integrand is symmetric under permutations
of $\{q^{(23)}_1,\dots,q^{(23)}_{n_{23}}\}$. We can choose to integrate only over the subspace for which $|q^{(23)}_1|\geq |q^{(23)}_j|$
for $j=2,\dots,n_{23}$, and compensate by multiplying with a factor $n_{23}$, i.e. multiply the integrand by
\be
  n_{23}\prod_{j=2}^{n_{23}}\Theta(|\bq_{23}-\bp_{23}|-|q^{(23)}_j|)\, ,
\ee
where $\bp_{23}\equiv \sum_{j=2}^{n_{23}} q^{(23)}_j$.
This guarantees that the factor $P_0(|\bq_{23}-\bp_{23}|)$ contained in $dQ$ after eliminating $q^{(23)}_1\to \bq_{23}-\bp_{23}=-\bk_2-\bp_{23}$ with the help of one of the Dirac deltas
is never evaluated for $|\bq_{23}-\bp_{23}|\to 0$, i.e. no infrared singularities appear.
An analogous modification can be made for the integration over $\{q^{(31)}_1,\dots,q^{(31)}_{n_{31}}\}$ if $n_{31}\geq 2$.
Finally, in order to guarantee a cancellation of infrared singularities for $Q_i\to 0$, $i=1,\dots,L$ among the various contributions, it is furthermore necessary to completely
symmetrize the integrand with respect to all 
\be
\frac{L!}{n_1!n_2!n_3!(n_{23}-1)!(n_{31}-1)!}
\ee
possibilities to choose the sets of wavenumbers $\{q^{(1)}_1,\dots,q^{(1)}_{n_1}\}$, 
$\{q^{(2)}_1,\dots,q^{(2)}_{n_2}\}$, $\{q^{(3)}_1,\dots,q^{(3)}_{n_1}\}$, $\{q^{(23)}_2,\dots,q^{(23)}_{n_{23}}\}$, $\{q^{(31)}_2,\dots,q^{(31)}_{n_{31}}\}$
out of the $L$ wavenumbers $Q_1,\dots,Q_L$, and further with respect to all $2^L$ possibilities to substitute $Q_i\to -Q_i$. At one-loop $L=1$ there is only
a single wavevector $Q_1$, and symmetrization with respect to $Q_1\to -Q_1$ is sufficient.

If the case (a) is realized, there is a loop running around all three `blobs' of the bispectrum. At one-loop, this is only the case for $B_{222}$,
and the $3=L+1$ integration variables in $dQ$ can be taken to be $\{Q_1,Q_2,Q_3\}=\{\bq_{12},\bq_{23},\bq_{31}\}$. Two of them can be eliminated by help of the
Dirac deltas in \eqref{eq:dQ}. In order to ensure cancellation of infrared singularities at the integrand level, one has to choose to eliminate the wavevector
with the largest norm, and the middle one. This can be achieved by inserting unity in the form of
\be
  1 = \int d^3q \left[ \delta^{(3)}(\bq-\bq_{12})\Theta(|\bq_{23}|-|\bq_{12}|)\Theta(|\bq_{31}|-|\bq_{12}|) + {\rm cyclic}\right]\, .
\ee
After eliminating the integrations over $\{\bq_{12},\bq_{23},\bq_{31}\}$ with the help of the Dirac deltas in the equation above and in \eqref{eq:dQ},
the remaining integration variable $\bq$ can be identified with $Q_1$. Finally, the integrand should be symmetrized with respect to $Q_1\to -Q_1$.
For $L>1$, one proceeds analogously to above, but in addition one has to symmetrize over all possibilities to choose $\bq$ from $Q_1,\dots,Q_L$.
If $n_{23}\geq 2$, one in addition performs the modifications analogous to case (b) described above, and similarly if $n_{12}\geq 2$ or $n_{31}\geq 2$.
For $L\geq 3$, it is in addition necessary to symmetrize over all possibilities to associate the $L-1$ remaining integration variables (i.e. after choosing $\bq$) to the internal
wavenumbers.

%%%%%%%%%%%%%%%%%%%%%%%%%%%%%%%%%%%%%%%%%%%%%%%%%%%%%%%%%%%%%%%%%%%%%%%%%%%%%%%%


\begin{thebibliography}{99}
%%%%%%%%%%%%%%%%%%%%%%%%%%%%%%%%%%%%%%%%%%%%%%%%%%%%%%%%%%%%%%%%%%%%%%%%%%%%%%%%

%\cite{Amendola:2012ys}
\bibitem{Amendola:2012ys}
  L.~Amendola {\it et al.} [Euclid Theory Working Group],
  %``Cosmology and fundamental physics with the Euclid satellite,''
  Living Rev.\ Rel.\  {\bf 16} (2013) 6
%  doi:10.12942/lrr-2013-6
  [arXiv:1206.1225 [astro-ph.CO]].
  %%CITATION = doi:10.12942/lrr-2013-6;%%

%\cite{Abell:2009aa}
\bibitem{Abell:2009aa}
  P.~A.~Abell {\it et al.} [LSST Science and LSST Project Collaborations],
  %``LSST Science Book, Version 2.0,''
  arXiv:0912.0201 [astro-ph.IM].
  %%CITATION = ARXIV:0912.0201;%%


%\cite{Levi:2013gra}
\bibitem{Levi:2013gra}
  M.~Levi {\it et al.} [DESI Collaboration],
  %``The DESI Experiment, a whitepaper for Snowmass 2013,''
  arXiv:1308.0847 [astro-ph.CO].
  %%CITATION = ARXIV:1308.0847;%%

%\cite{Dawson:2015wdb}
\bibitem{Dawson:2015wdb}
  K.~S.~Dawson {\it et al.},
  %``The SDSS-IV extended Baryon Oscillation Spectroscopic Survey: Overview and Early Data,''
  Astron.\ J.\  {\bf 151} (2016) 44
%  doi:10.3847/0004-6256/151/2/44
  [arXiv:1508.04473 [astro-ph.CO]].
  %%CITATION = doi:10.3847/0004-6256/151/2/44;%%

%\cite{Ata:2017dya}
\bibitem{Ata:2017dya}
  M.~Ata {\it et al.},
  %``The clustering of the SDSS-IV extended Baryon Oscillation Spectroscopic Survey DR14 quasar sample: first measurement of baryon acoustic oscillations between redshift 0.8 and 2.2,''
  Mon.\ Not.\ Roy.\ Astron.\ Soc.\  {\bf 473} (2018) no.4,  4773
  doi:10.1093/mnras/stx2630
  [arXiv:1705.06373 [astro-ph.CO]].
  %%CITATION = doi:10.1093/mnras/stx2630;%%
  %100 citations counted in INSPIRE as of 29 Apr 2019 

%\cite{Kuhlen:2012ft}
\bibitem{Kuhlen:2012ft}
  M.~Kuhlen, M.~Vogelsberger and R.~Angulo,
  %``Numerical Simulations of the Dark Universe: State of the Art and the Next Decade,''
  Phys.\ Dark Univ.\  {\bf 1} (2012) 50
%  doi:10.1016/j.dark.2012.10.002
  [arXiv:1209.5745 [astro-ph.CO]].


%\cite{Vogelsberger:2014dza}
\bibitem{Vogelsberger:2014dza}
  M.~Vogelsberger {\it et al.},
  %``Introducing the Illustris Project: Simulating the coevolution of dark and visible matter in the Universe,''
  Mon.\ Not.\ Roy.\ Astron.\ Soc.\  {\bf 444} (2014) no.2,  1518
%  doi:10.1093/mnras/stu1536
  [arXiv:1405.2921 [astro-ph.CO]].
  %%CITATION = doi:10.1093/mnras/stu1536;%%

%\cite{Schneider:2015yka}
\bibitem{Schneider:2015yka}
  A.~Schneider {\it et al.},
  %``Matter power spectrum and the challenge of percent accuracy,''
  JCAP {\bf 1604} (2016) no.04,  047
%  doi:10.1088/1475-7516/2016/04/047
  [arXiv:1503.05920 [astro-ph.CO]].
  %%CITATION = doi:10.1088/1475-7516/2016/04/047;%%

\bibitem{bernardeau0}
  F.~Bernardeau, S.~Colombi, E.~Gaztanaga and R.~Scoccimarro,
  %``Large scale structure of the universe and cosmological perturbation theory,''
  Phys.\ Rept.\  {\bf 367} (2002) 1
  [astro-ph/0112551].
  %%CITATION = ASTRO-PH/0112551;%%

\bibitem{bernardeau}
  F.~Bernardeau,
  %``The evolution of the large-scale structure of the universe: beyond the linear regime,''
  arXiv:1311.2724 [astro-ph.CO].
  %%CITATION = ARXIV:1311.2724;%%

\bibitem{CrSc1}
  M.~Crocce and R.~Scoccimarro,
  %``Renormalized Cosmological Perturbation Theory,''
  Phys.\ Rev.\  D {\bf 73} (2006) 063519
  [arXiv:astro-ph/0509418].
  %%CITATION = PHRVA,D73,063519;%%

%\cite{Audren:2011ne}
\bibitem{Audren:2011ne}
  B.~Audren and J.~Lesgourgues,
  %``Non-linear matter power spectrum from Time Renormalisation Group: efficient computation and comparison with one-loop,''
  JCAP {\bf 1110} (2011) 037
  [arXiv:1106.2607 [astro-ph.CO]].
  %%CITATION = ARXIV:1106.2607;%%


%\cite{Carlson:2009it}
\bibitem{Carlson:2009it}
  J.~Carlson, M.~White and N.~Padmanabhan,
  %``A critical look at cosmological perturbation theory techniques,''
  Phys.\ Rev.\ D {\bf 80} (2009) 043531
%  doi:10.1103/PhysRevD.80.043531
  [arXiv:0905.0479 [astro-ph.CO]].
  %%CITATION = doi:10.1103/PhysRevD.80.043531;%%
  %116 citations counted in INSPIRE as of 01 Jul 2016

%\cite{Taruya:2012ut}
\bibitem{Taruya:2012ut}
  A.~Taruya, F.~Bernardeau, T.~Nishimichi and S.~Codis,
  %``RegPT: Direct and fast calculation of regularized cosmological power spectrum at two-loop order,''
  Phys.\ Rev.\ D {\bf 86} (2012) 103528
%  doi:10.1103/PhysRevD.86.103528
  [arXiv:1208.1191 [astro-ph.CO]].
  %%CITATION = doi:10.1103/PhysRevD.86.103528;%%

%\cite{Scoccimarro:1995if}
\bibitem{Scoccimarro:1995if}
  R.~Scoccimarro and J.~Frieman,
  %``Loop corrections in nonlinear cosmological perturbation theory,''
  Astrophys.\ J.\ Suppl.\  {\bf 105} (1996) 37
%  doi:10.1086/192306
  [astro-ph/9509047].
  %%CITATION = doi:10.1086/192306;%%

  
 \bibitem{CrSc2}
  M.~Crocce and R.~Scoccimarro,
 %``Memory of Initial Conditions in Gravitational Clustering,''
  Phys.\ Rev.\  D {\bf 73} (2006) 063520
  [arXiv:astro-ph/0509419].
  %%CITATION = PHRVA,D73,063520;%%
 
 %\cite{Bernardeau:2011vy}
\bibitem{Bernardeau:2011vy}
  F.~Bernardeau, N.~Van de Rijt and F.~Vernizzi,
  %``Resummed propagators in multi-component cosmic fluids with the eikonal approximation,''
  Phys.\ Rev.\ D {\bf 85} (2012) 063509
%  doi:10.1103/PhysRevD.85.063509
  [arXiv:1109.3400 [astro-ph.CO]].
  %%CITATION = doi:10.1103/PhysRevD.85.063509;%%


\bibitem{nonlinear}
  D.~Blas, M.~Garny and T.~Konstandin,
  %``On the non-linear scale of   cosmological perturbation theory,''
  JCAP {\bf 1309} (2013) 024
  [arXiv:1304.1546 [astro-ph.CO]].
  %%CITATION = ARXIV:1304.1546;%% 
 
%\cite{Crocce:2007dt}
\bibitem{Crocce:2007dt}
  M.~Crocce and R.~Scoccimarro,
  %``Nonlinear Evolution of Baryon Acoustic Oscillations,''
  Phys.\ Rev.\ D {\bf 77} (2008) 023533
  [arXiv:0704.2783 [astro-ph]].
  %%CITATION = ARXIV:0704.2783;%%

  
  %\cite{Smith:2007gi}
\bibitem{Smith:2007gi}
  R.~E.~Smith, R.~Scoccimarro and R.~K.~Sheth,
  %``Eppur Si Muove: On The Motion of the Acoustic Peak in the Correlation Function,''
  Phys.\ Rev.\ D {\bf 77} (2008) 043525
%  doi:10.1103/PhysRevD.77.043525
  [astro-ph/0703620 [ASTRO-PH]].
  %%CITATION = doi:10.1103/PhysRevD.77.043525;%%

%\cite{Sherwin:2012nh}
\bibitem{Sherwin:2012nh}
  B.~D.~Sherwin and M.~Zaldarriaga,
  %``The Shift of the Baryon Acoustic Oscillation Scale: A Simple Physical Picture,''
  Phys.\ Rev.\ D {\bf 85} (2012) 103523
%  doi:10.1103/PhysRevD.85.103523
  [arXiv:1202.3998 [astro-ph.CO]].
  %%CITATION = doi:10.1103/PhysRevD.85.103523;%%

%\cite{Senatore:2014via}
\bibitem{Senatore:2014via}
  L.~Senatore and M.~Zaldarriaga,
  %``The IR-resummed Effective Field Theory of Large Scale Structures,''
  JCAP {\bf 1502} (2015) no.02,  013
%  doi:10.1088/1475-7516/2015/02/013
  [arXiv:1404.5954 [astro-ph.CO]].
  %%CITATION = doi:10.1088/1475-7516/2015/02/013;%%

%\cite{Blas:2015qsi}  
\bibitem{Blas:2015qsi}
  D.~Blas, M.~Garny, M.~M.~Ivanov and S.~Sibiryakov,
  %``Time-Sliced Perturbation Theory for Large Scale Structure I: General Formalism,''
  JCAP {\bf 1607} (2016) no.07,  052
%  doi:10.1088/1475-7516/2016/07/052
  [arXiv:1512.05807 [astro-ph.CO]].
  %%CITATION = doi:10.1088/1475-7516/2016/07/052;%%



%\cite{Blas:2016sfa}
\bibitem{Blas:2016sfa}
  D.~Blas, M.~Garny, M.~M.~Ivanov and S.~Sibiryakov,
  %``Time-Sliced Perturbation Theory II: Baryon Acoustic Oscillations and Infrared Resummation,''
  JCAP {\bf 1607} (2016) no.07,  028
%  doi:10.1088/1475-7516/2016/07/028
  [arXiv:1605.02149 [astro-ph.CO]].
  %%CITATION = doi:10.1088/1475-7516/2016/07/028;%%

 \bibitem{Blas1}
  D.~Blas, M.~Garny and T.~Konstandin,
  %``Cosmological perturbation theory at three-loop order,''
  JCAP {\bf 1401} (2014) 010
  [arXiv:1309.3308 [astro-ph.CO]].
  %%CITATION = ARXIV:1309.3308;%%
 

  %\cite{Baumann:2010tm}
\bibitem{Baumann:2010tm}
  D.~Baumann, A.~Nicolis, L.~Senatore and M.~Zaldarriaga,
  %``Cosmological Non-Linearities as an Effective Fluid,''
  JCAP {\bf 1207} (2012) 051
  [arXiv:1004.2488 [astro-ph.CO]].
  %%CITATION = ARXIV:1004.2488;%%


%\cite{Carrasco:2012cv}
\bibitem{Carrasco:2012cv}
  J.~J.~M.~Carrasco, M.~P.~Hertzberg and L.~Senatore,
  %``The Effective Field Theory of Cosmological Large Scale Structures,''
  JHEP {\bf 1209} (2012) 082
  [arXiv:1206.2926 [astro-ph.CO]].
  %%CITATION = ARXIV:1206.2926;%%
  
  %\cite{Porto:2013qua}
\bibitem{Porto:2013qua} 
  R.~A.~Porto, L.~Senatore and M.~Zaldarriaga,
  %``The Lagrangian-space Effective Field Theory of Large Scale Structures,''
  JCAP {\bf 1405}, 022 (2014)
  [arXiv:1311.2168 [astro-ph.CO]].
  %%CITATION = ARXIV:1311.2168;%%

%\cite{Baldauf:2014qfa}
\bibitem{Baldauf:2014qfa}
  T.~Baldauf, L.~Mercolli, M.~Mirbabayi and E.~Pajer,
  %``The Bispectrum in the Effective Field Theory of Large Scale Structure,''
  JCAP {\bf 1505} (2015) no.05,  007
%  doi:10.1088/1475-7516/2015/05/007
  [arXiv:1406.4135 [astro-ph.CO]].
  %%CITATION = doi:10.1088/1475-7516/2015/05/007;%%
  %73 citations counted in INSPIRE as of 22 Feb 2019

%\cite{Angulo:2014tfa}
\bibitem{Angulo:2014tfa}
  R.~E.~Angulo, S.~Foreman, M.~Schmittfull and L.~Senatore,
  %``The One-Loop Matter Bispectrum in the Effective Field Theory of Large Scale Structures,''
  JCAP {\bf 1510} (2015) no.10,  039
%  doi:10.1088/1475-7516/2015/10/039
  [arXiv:1406.4143 [astro-ph.CO]].
  %%CITATION = doi:10.1088/1475-7516/2015/10/039;%%
  %58 citations counted in INSPIRE as of 22 Feb 2019
  
 %\cite{Foreman:2015lca}
\bibitem{Foreman:2015lca}
  S.~Foreman, H.~Perrier and L.~Senatore,
  %``Precision Comparison of the Power Spectrum in the EFTofLSS with Simulations,''
  JCAP {\bf 1605} (2016) no.05,  027
%  doi:10.1088/1475-7516/2016/05/027
  [arXiv:1507.05326 [astro-ph.CO]].
  %%CITATION = doi:10.1088/1475-7516/2016/05/027;%%


%\cite{Baldauf:2015aha}
\bibitem{Baldauf:2015aha}
  T.~Baldauf, L.~Mercolli and M.~Zaldarriaga,
  %``Effective field theory of large scale structure at two loops: The apparent scale dependence of the speed of sound,''
  Phys.\ Rev.\ D {\bf 92} (2015) no.12,  123007
%  doi:10.1103/PhysRevD.92.123007
  [arXiv:1507.02256 [astro-ph.CO]].
  %%CITATION = doi:10.1103/PhysRevD.92.123007;%%


%\cite{Assassi:2015jqa}
\bibitem{Assassi:2015jqa}
  V.~Assassi, D.~Baumann, E.~Pajer, Y.~Welling and D.~van der Woude,
  %``Effective theory of large-scale structure with primordial non-Gaussianity,''
  JCAP {\bf 1511} (2015) 024
%  doi:10.1088/1475-7516/2015/11/024
  [arXiv:1505.06668 [astro-ph.CO]].
  %%CITATION = doi:10.1088/1475-7516/2015/11/024;%%


%\cite{Abolhasani:2015mra}
\bibitem{Abolhasani:2015mra}
  A.~A.~Abolhasani, M.~Mirbabayi and E.~Pajer,
  %``Systematic Renormalization of the Effective Theory of Large Scale Structure,''
  JCAP {\bf 1605} (2016) no.05,  063
%  doi:10.1088/1475-7516/2016/05/063
  [arXiv:1509.07886 [hep-th]].
  %%CITATION = doi:10.1088/1475-7516/2016/05/063;%%


%\cite{Fuhrer:2015cia}
\bibitem{Fuhrer:2015cia}
  F.~F\"{u}hrer and G.~Rigopoulos,
  %``Renormalizing a Viscous Fluid Model for Large Scale Structure Formation,''
  JCAP {\bf 1602} (2016) no.02,  032
%  doi:10.1088/1475-7516/2016/02/032
  [arXiv:1509.03073 [astro-ph.CO]].
  %%CITATION = doi:10.1088/1475-7516/2016/02/032;%%

%\cite{Chudaykin:2019ock}
\bibitem{Chudaykin:2019ock}
  A.~Chudaykin and M.~M.~Ivanov,
  %``Measuring neutrino masses with large-scale structure: Euclid forecast with controlled theoretical error,''
  arXiv:1907.06666 [astro-ph.CO].
  %%CITATION = ARXIV:1907.06666;%%
  
\bibitem{MSR73}
P.C.~Martin, E.D.~Siggia and H.A.~Rose, 
%``Statistical Dynamics of Classical Systems,''
Phys.Rev. A {\bf 8} (1973) 423.

\bibitem{Max1}    
S.~Matarrese and M.~Pietroni,
  %``Resumming Cosmic Perturbations,''
  JCAP {\bf 0706} (2007) 026
  [arXiv:astro-ph/0703563].
  %%CITATION = JCAPA,0706,026;%%

\bibitem{viscousdm}
%\cite{Blas:2015tla}
%\bibitem{Blas:2015tla}
  D.~Blas, S.~Floerchinger, M.~Garny, N.~Tetradis and U.~A.~Wiedemann,
  %``Large scale structure from viscous dark matter,''
  JCAP {\bf 1511} (2015) 049
%  doi:10.1088/1475-7516/2015/11/049
  [arXiv:1507.06665 [astro-ph.CO]].
  %%CITATION = doi:10.1088/1475-7516/2015/11/049;%%


%\cite{Floerchinger:2016hja}
\bibitem{Floerchinger:2016hja}
  S.~Floerchinger, M.~Garny, N.~Tetradis and U.~A.~Wiedemann,
  %``Renormalization-group flow of the effective action of cosmological large-scale structures,''
  JCAP {\bf 1701} (2017) no.01,  048
%  doi:10.1088/1475-7516/2017/01/048
  [arXiv:1607.03453 [astro-ph.CO]].
  %%CITATION = doi:10.1088/1475-7516/2017/01/048;%%

  
%\cite{Berges:2000ew}
\bibitem{Berges:2000ew}
  J.~Berges, N.~Tetradis and C.~Wetterich,
  %``Nonperturbative renormalization flow in quantum field theory and statistical physics,''
  Phys.\ Rept.\  {\bf 363} (2002) 223
%  doi:10.1016/S0370-1573(01)00098-9
  [hep-ph/0005122].
  %%CITATION = doi:10.1016/S0370-1573(01)00098-9;%%


%\cite{Pietroni:2011iz}
\bibitem{Pietroni:2011iz}
  M.~Pietroni, G.~Mangano, N.~Saviano and M.~Viel,
  %``Coarse-Grained Cosmological Perturbation Theory,''
  JCAP {\bf 1201} (2012) 019
%  doi:10.1088/1475-7516/2012/01/019
  [arXiv:1108.5203 [astro-ph.CO]].
  %%CITATION = doi:10.1088/1475-7516/2012/01/019;%%


%\cite{Manzotti:2014loa}
\bibitem{Manzotti:2014loa}
  A.~Manzotti, M.~Peloso, M.~Pietroni, M.~Viel and F.~Villaescusa-Navarro,
  %``A coarse grained perturbation theory for the Large Scale Structure, with cosmology and time independence in the UV,''
  JCAP {\bf 1409} (2014) no.09,  047
%  doi:10.1088/1475-7516/2014/09/047
  [arXiv:1407.1342 [astro-ph.CO]].
  %%CITATION = doi:10.1088/1475-7516/2014/09/047;%%


\bibitem{goroff}
    M.~H.~Goroff, B.~Grinstein, S.~J.~Rey and M.~B.~Wise,
    %``Coupling of Modes of Cosmological Mass Density Fluctuations,''
    Astrophys.\ J.\  {\bf 311} (1986) 6.
 %   doi:10.1086/164749
    %%CITATION = doi:10.1086/164749;%%
    
    
%\cite{Blas:2011rf}
\bibitem{Blas:2011rf}
  D.~Blas, J.~Lesgourgues and T.~Tram,
  %``The Cosmic Linear Anisotropy Solving System (CLASS) II: Approximation schemes,''
  JCAP {\bf 1107} (2011) 034
%  doi:10.1088/1475-7516/2011/07/034
  [arXiv:1104.2933 [astro-ph.CO]].
  %%CITATION = doi:10.1088/1475-7516/2011/07/034;%%


%\cite{McDonald:2009hs}
\bibitem{McDonald:2009hs}
  P.~McDonald,
  %``How to generate a significant effective temperature for cold dark matter, from first principles,''
  JCAP {\bf 1104} (2011) 032
%  doi:10.1088/1475-7516/2011/04/032
  [arXiv:0910.1002 [astro-ph.CO]].
  %%CITATION = doi:10.1088/1475-7516/2011/04/032;%%
  %19 citations counted in INSPIRE as of 06 Apr 2019

%\cite{Erschfeld:2018zqg}
\bibitem{Erschfeld:2018zqg}
  A.~Erschfeld and S.~Floerchinger,
  %``Evolution of dark matter velocity dispersion,''
  JCAP {\bf 1906} (2019) no.06,  039
  doi:10.1088/1475-7516/2019/06/039
  [arXiv:1812.06891 [astro-ph.CO]].
  %%CITATION = doi:10.1088/1475-7516/2019/06/039;%%
  %1 citations counted in INSPIRE as of 23 Jul 2019

%\cite{Lazanu:2018yae}
\bibitem{Lazanu:2018yae}
  A.~Lazanu and M.~Liguori,
  %``The two and three-loop matter bispectrum in perturbation theories,''
  JCAP {\bf 1804} (2018) no.04,  055
  doi:10.1088/1475-7516/2018/04/055
  [arXiv:1803.03184 [astro-ph.CO]].
  %%CITATION = doi:10.1088/1475-7516/2018/04/055;%%
  %2 citations counted in INSPIRE as of 23 Jul 2019

\end{thebibliography}
\end{document}